\documentclass[aps,pra,twocolumn,superscriptaddress,showkeys,10pt]{revtex4-1}
\usepackage{amssymb, amsmath,color,mciteplus,graphicx,subfigure}
\usepackage{calc,epsfig,epstopdf,color,mciteplus,bm,mathrsfs}
\usepackage{times}
\newcommand{\abs}[1]{\left| {#1} \right|}
\newcommand{\dd}{d}

\newcommand{\ket}[1]{\ensuremath{\left|#1\right\rangle}}

\begin{document}

\title{Topological photonics on superconducting quantum circuits with parametric couplings}

\author{Zheng-Yuan Xue}\email{zyxue83@163.com}
\affiliation{Guangdong Provincial Key Laboratory of Quantum Engineering and Quantum Materials,
and School of Physics\\ and Telecommunication Engineering, South China Normal University, Guangzhou 510006, China}
\affiliation{Guangdong-Hong Kong Joint Laboratory of Quantum Matter, and Frontier Research Institute for Physics,\\ South China Normal University, Guangzhou 510006, China}

\author{Yong Hu}\email{yonghu@hust.edu.cn}

\affiliation{School of Physics, Huazhong University of Science and Technology, Wuhan 430074, China}

\date{\today}

\begin{abstract}

Topological phases of matter is an exotic phenomena in modern condense matter physics, which has attracted much attention due to the unique boundary states and transport properties. Recently, this topological concept in electronic materials has been exploited in many other fields of physics. Motivated by designing and controlling the behavior of electromagnetic waves, in optical, microwave, and sound frequencies, topological photonics emerges as a rapid growing  research field. Due to the flexibility and diversity of superconducting quantum circuits system, it is an promising platform to realize exotic topological phases of matter  and to probe and explore topologically-protected effects in new ways. Here, we review theoretical and experimental advances of topological photonics on superconducting quantum circuits via the experimentally demonstrated parametric tunable coupling techniques, including using of the superconducting transmission line resonator, superconducting qubits, and the coupled system of them. On superconducting circuits, the flexible interactions and intrinsic nonlinearity  making topological photonics in this system not only a simple photonic analog of topological effects for novel devices, but also a realm of exotic but less-explored fundamental physics.

\end{abstract}


\maketitle

\section{Introduction}

Topology considers the geometric properties of objects that can be preserved under continuous deformations, such as stretching and bending, and quantities remaining invariant under such continuous deformations are denoted as topological invariants \cite{ArmstrongTopologyBook,NakaharaGeomeryBook,FrankelGeometryBook}. For instance, a two-dimensional (2D) closed surface embedded in 3D space has the topological invariant genus which counts the number of the holes within the surface. This topological invariant does not respond to small perturbations as long as the holes are not created by tearing or removed by gluing. Objects which look very different can be equivalent in the sense of topology, and objects with different topologies can be classified into different equivalent classes by their topological invariants. For instance,  a sphere and a spoon are equivalent because both of their genus are zero, while a torus and a coffee cup are also topological equivalent as their genus are one.


Historically, the topological concept combined with condensed matter physics through the discovery of the quantum Hall effect (QHE) \cite{KlitzingQHE1980PRL}, in which the quantized  Hall conductance of 2D electronic materials can be interpreted by the topological  Chern number of the filled Landau levels \cite{TKNN1982PRL}. This exotic phase of matters cannot be explained by the conventional spontaneous symmetry-breaking mechanism.  Instead, it should be understood within the framework of the newly-developed topological band theory (TBT) \cite{HasanKane2010RMP, QiZhang2011RMP, BernevigTIBook2013, Bansil2016RMP}. A direct consequence of a nontrivial topological band structure is the emergence of edge state modes (ESMs), i.e. the bulk-edge correspondence \cite{HatsugaiBEC1993PRL, HatsugaiBEC1993PRB, QiXLBEC2006PRB, pnas}. The existence of the ESMs can be explained by the following simple argument \cite{LaughlinBEC1981PRB}: we put two insulators with different topological band structures together to share a boundary, and away from the boundary the two insulators extend to infinity. Then, these two insulators can not be connected trivially at the boundary region, and thus  a topological phase transition has to happen  somewhere when the band gaps are closed and reopened, i.e. there must have ESMs crossing the band gap. The presence of the gapless ESMs can thus be treated as an unambiguous signature of the topological non-trivial band structure of the bulk. The gapless spectra of the ESMs are topologically protected as their existence is guaranteed by the difference of the topologies of the bulk materials on the two sides, and their propagation is robust against perturbations in the materials including disorder and defects.

As TBT mainly considers the band topologies of Hermitian electronic systems in the single-particle picture, the obtained results can be directly applied to interaction-free bosonic systems, leading to the emerging research field of topological photonics \cite{LuTPReviewNatPhoton2014, TopoPhoton2019RMP}. During the past few years, there are numerous theoretical and experimental explorations of the photonic analogue of the QHE in various artificial photonic metamaterials \cite{HaldaneTPPRL,TPNature2009,FanSHDMNP2012,HafeziNatPhoton2013}. The motivation lies in both the sides of experimental application and fundamental physics. On  integrated optical circuits,  the back-reflection induced by disorder and defects is a major hindrance of efficient  information transmission. Therefore, it is expected that the unidirectional ESMs, if synthesized, can be used to transmit electromagnetic waves without the back-reflection even in the presence of large disorder. This closely mimics the topologically protected quantized Hall conductance of 2D electronic materials, and such ideal transport may offer novel functionalities for photonic systems with topological immunity to fabrication errors or environmental changes. On the other hand, topological photonics also brings new physics of fundamental importance. As the band topology of spinless particles remains trivial as long as the time-reversal symmetry is preserved,  effective magnetic fields have to be synthesized for the charge-neutral photons in the metamaterials \cite{FanSHDMNP2012,HafeziNatPhoton2013,HafeziNatPhys2011}, which is represented by the nontrivial hopping phases on the lattice through Peierel's substitution. In condensed matter physics,   the electrons' magnetic lengths of the the current-achievable strong magnetic field are still far more larger than the lattice spacing \cite{GoerbigReview,MagneticField}. Meanwhile, based on the fabrication and manipulation of artificial photonic metamaterials, it is expected to obtain arbitrarily large Aharonov-Bohm phase for hopping photons in a loop containing only few unit-cells, indicating that the synthetic magnetic field for photons can be  several orders larger than those for electronic materials. In addition, it should be noticed that the photonic systems is naturally non-equilibrium and bosonic, exhibiting neither chemical potential nor fermionic statistics. This is in stark contrast to the conventional electronic materials. Therefore, from the above two points of view, topological photonics is not only the simple photonic analog of known topological phases of matter, but also a realm of exotic but less-explored fundamental physics.

Here, we focus on reviewing experimental and theoretical advances of topological photonics on superconducting quantum circuits (SQC) \cite{MakhlinQC2001RMP, sqc2, JQYouReview, sqc4, LiuYXReview2017PR,CQED2021}, including using of the superconducting qubits, superconducting transmission line resonator (TLR) and the coupled system of them. SQC can be regarded as the realization of atom-photon interaction \cite{HarocheCQEDReview} in on-chip Josephson junction based mesoscopic electronic circuit. It uses the superconducting TLRs \cite{YaleCQEDPRA2004, WallraffNature2004} to replace the standing-wave optical cavities and the superconducting qubits \cite{MakhlinQC2001RMP, KochTransmonPRA2007} to substitute the atoms.  Historically, this on-chip superconducting architecture is proposed as a candidate platform for quantum computation. While borrowed a variety of ideas from atomic cavity QED \cite{HarocheCQEDReview} in its early development, SQC has now become an independent research field due to its unprecedented advantages including the strong nonlinearity at the single photon level, the long coherence times, the flexibility in circuit design, the detailed control of atom-photon interaction at the quantum level, and the scalability based on current microelectronic technology \cite{HouckReview2012NP, SchmidtReview2013ADP, CarusottoReview2020NP}. Recently, it has been realized that the above mentioned merits are also essential for the synthesization of photonic metamaterial in the microwave regime and the quantum simulation of various complicated many-body effects. For the purpose of quantum simulation,  individual SQC elements such as superconducting TLRs  or  qubits play the role of the bosons and/or fermions in lattice models, and the inter-site coupling is established through the connection of the SQC elements via various kinds of couplers. From this point of view, a coupler which can support both tunable coupling strength and hopping phase is highly desirable. The former can be exploited to the investigation of the competition between the inter-site hopping and the on-site Hubbard interaction \cite{HartmannCoupledCavity2006NP, CarusottoFermionizedPhoton2009PRL, CarusottoQFLRMP2013}, while the latter, being equivalent to the artificial gauge field for the photons on the lattice \cite{GoldmanGauge2014RPP,TopoPhoton2019RMP,GoldmanGauge2018}, is required to realize non-trivial topological effect, which is our central topic here.

The research of synthesizing gauge fields in artificial models was originated in the  ultracold atomic systems \cite{SpielmanReview, GoldmanGauge2014RPP, TopoAtom2019RMP}, while SQC system takes the naturally advantages of individual addressing and  \textit{in situ} tunability of circuit parameters \cite{JQYouReview,HouckReview2012NP,SchmidtReview2013ADP}.  
The other distinct merits of quantum simulation topological phases with SQC are the following aspects. First, the topological state of matters are highly related with the geometry of the lattice. Meanwhile, by taking the advantage of arbitrary wiring,  SQC can be used to implement lattice configurations which are difficult for other simulators. Second, the effective strong interaction between photons can be synthesized in a tunable way via many mechanisms, including the electromagnetically induced transparency \cite{HartmannCoupledCavity2006NP, HuYongCrossKerr2011PRA}, Jaynes-Cummings-Hubbard nonlinearity \cite{GreenTreeJCHNP2006}, as well as nonlinear Josephson coupling \cite{BourassaSQUIDCouplingPRA2012, JinJSPRL2013, LiebSQUIDCouplingNJP2012}. This can be understood by the fact that a superconducting qubit can be treated as a nonlinear resonator in the microwave regime \cite{Noh}, and thus the topological photonics on SQC can be directly extended to the nonlinear regime, i.e., nonlinear topological photonics \cite{nonlinearSSHsimu}. In particular, the Hamiltonian of a qubit chain can be described by the Bose-Hubbard model, where the nonlinearities can naturally introduce photonic interactions \cite{RoushanChiral2018Science}, and thus interaction induced topological photonics can be possible. As the number of lattice sites grows, the quantum dynamics quickly become intractable even for classical super-computers. Therefore, combining the strong and tunable photonic  interaction with the synthetic gauge fields make the SQC system promising in realizing bosonic fractional QHE \cite{UmucalilarFQHEPRL2012,HaywardFQHE2012PRL} and nontrivial topological ESMs for photons in the microwave regime \cite{ImamogluMajoranaPRL2012}. Such method can offer facilities in the study of the competition effect between synthetic gauge fields, photon hopping, and Hubbard repulsion. The third point is that, as the control of superconducting devices and their interaction is now very precise and diverse, it is possible to engineer the gain and loss of a system. In this sense, SQC can be exploited to investigate the  non-Hermitian induced topological phases which exhibit drastic difference from their Hermitian counterpart.

\section{Introduction to  TBT}
In this section, we briefly review the basic concepts of  TBT. Currently there have already been many excellent reviews and tutorials on this subject, see, e.g., Ref. \cite{BernevigTIBook2013} for the pedagogical tutorial of TBT and Ref.  \cite{TopoPhoton2019RMP} for the summary of recent advances in topological photonics.

\subsection{The topological invariant}
\begin{figure}[tbp]
\begin{center}
\includegraphics[width=0.48\textwidth]{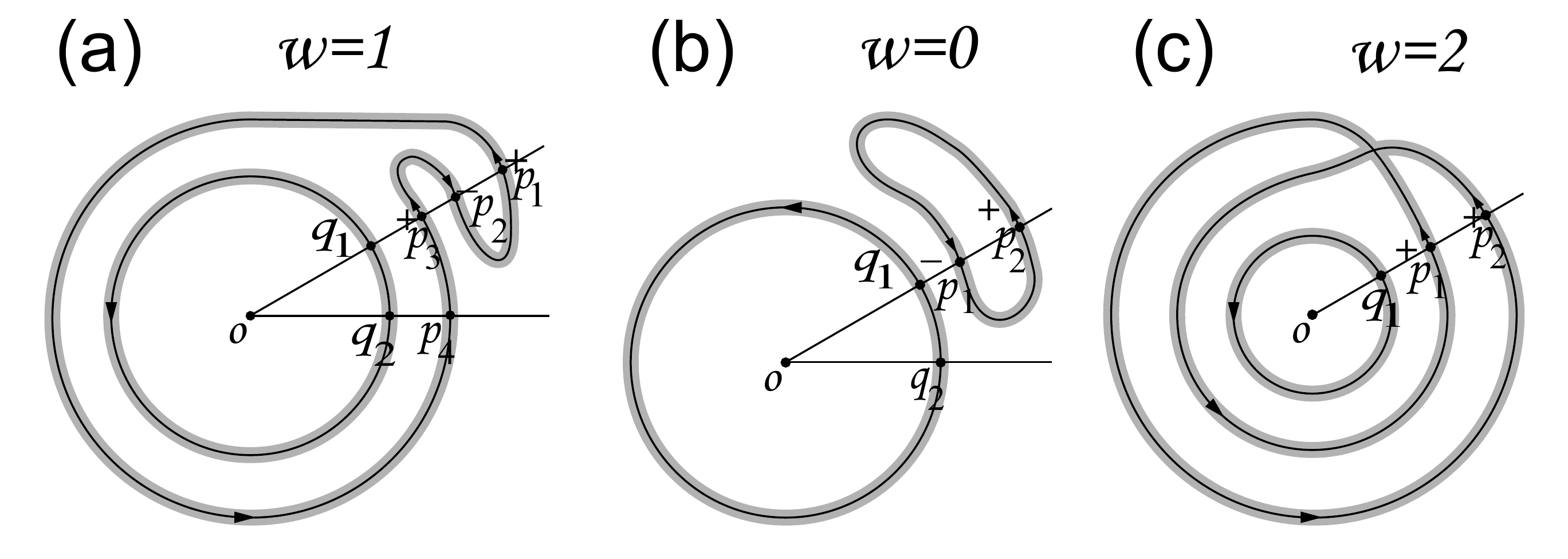}
\end{center}
\caption{Winding number $w$ of 1D closed loops, which  counts the algebraic times an image point on $S^1$ is radically mapped, or equivalently the times a loop wraps around the origin. For the loops sketched, $w$ takes values (a) $w=1$, (b) $w=0$, and (c) $w=2$, and exhibit topological robustness against the small variation of the loops.}
\label{Fig WindingNumber}
\end{figure}

Most topological invariants being involved in this review can be understood in the framework of degree of map \cite{FrankelGeometryBook}: For a smooth map between two closed oriented manifolds with the same dimension, an integer can be defined to characterize the algebraic times an image point is covered. To illustrate this concept, let us take the winding number $w$ of a one-dimension (1D) closed loop $M$ parameterized by $\phi \in \left[ 0, 2\pi\right)$ as an example, see Section \ref{TBTin2D} for the two-dimension (2D) case. As shown in Fig. \ref{Fig WindingNumber}, we consider the radical map $f$ from $M$ to the 1D unit circle $S^1$ parameterized by the azimuthal angle $\theta$. The rule of $f$ is that we move each point of $M$ radically until it strikes $S^1$. Since the variation of $f$ can be equivalently realized by keeping the radical rule but changing the shape of $M$, the degree of map of $f$ in this situation is recognized as the winding number of of the loop $M$, which is the Jacobian integration of $f$ over $M$,
\begin{equation}
	\label{Eqn DegreeofMapIntegration}
	w=\frac{1}{2\pi}  \int_{\phi=0}^{\phi=2\pi} \frac{d \theta}{d \phi} d {\phi} \in \mathbb{Z}.
	\end{equation}
Here the $1/2\pi$ factor indicates the length of the target manifold $S^1$. Clearly, $w$ measures the times that the target manifold $S^1$ is covered by the image of $M$ under $f$.  There is another equivalent discrete form of $w$. For the loop in Fig. \ref{Fig WindingNumber}(a), let us choose a point $q_1 \in S^1$ and count its inverse images $p_1$, $p_2$, and $p_3$. In this counting, every inverse image contribute $+1$ or $-1$, depending on the direction, i.e. the sign of the Jacobian of $f$, at that inverse image. Therefore, $p_1$ and $p_3$ contribute $+1$ each and $p_2$ contributes $-1$, resulting an overall $w=1$ (see also Figs. \ref{Fig WindingNumber}(b) and (c) for the $w=0$ and $w=2$ cases). Notice that $w$ does not depend on the choice of the image point: we can select another image point $q_2$, and its single inverse image $p_4$ also results in an overall $+1$ counting.

The topological nature of $w$ is manifested by the fact that it dose not alter under the slight deformation of $M$ as long as the deformation  does not hit the origin where the radical rule is ill-defined. We can check the other loops in Fig. \ref{Fig WindingNumber}(b) and (c), which have winding numbers $0$ and $2$, respectively. It is obvious that we can not change their winding numbers until we let some of the point on the loop cross the origin. Such topological robustness is intuitive: as shown in Eq. \eqref{Eqn DegreeofMapIntegration}, $w$ is a continuous and integer-valued functional of $M$. Therefore, under the continuous deformation of $M$, $w$ can not change continuously, but has to either remain invariant or change abruptly.

\subsection{TBT in 1D: SSH Model} \label{SSH}
\begin{figure}[tbp]
\centering
\includegraphics[width=8cm]{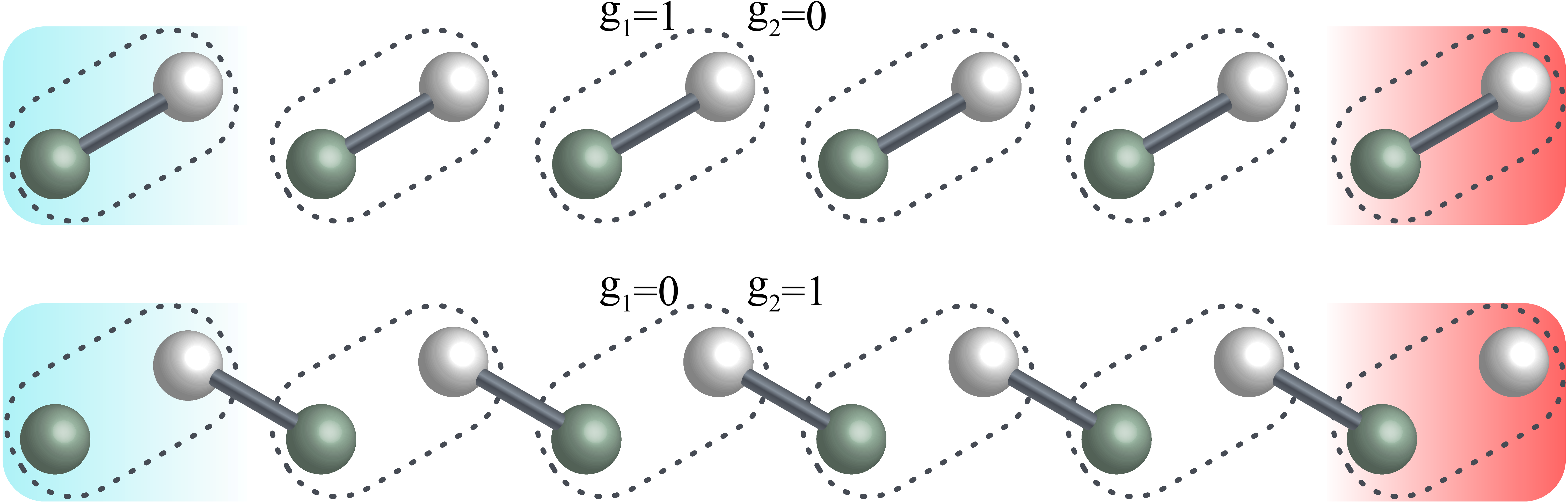}
\caption{\label{Fig SSH} Sketch of  the SSH model in the two topological trivial (a) and non-trivial (b) dimerization limits.}
\end{figure}

A prominent example of closed manifolds in physics is the first Brillouin zone  of periodic lattices,  and perhaps the most typical and simple model exhibiting nontrivial band topology is the 1D  Su-Schrieffer-Heeger (SSH) model describing polyacetylene chain formed by bonded CH groups \cite{SSH}. As shown in Fig. \ref{Fig SSH}, the SSH model describes electrons hopping on a chain with two sublattices and staggered hopping constants. The Hamiltonian takes the form
\begin{equation}
	\label{Eqn SSHHamiltonian}
	H_{\mathrm{SSH}}= \sum_{n} \left[ g_A a_{n,A}^{\dagger} a_{n,B} +g_B a_{n+1,A}^{\dagger} a_{nB} \right]+\mathrm{H.c.},
	\end{equation}
where $a_{n,\alpha}$ ($a_{n,\alpha}^{\dagger}$) is the annihilation (creation) operator of the $\alpha$th sublattice in the $n$th unit-cell, and $g_A$, $g_B$ denotes the intra- and inter-unit-cell hopping strength, respectively. In the reciprocal space, $H_{\mathrm{SSH}}$ has the form
	\begin{equation}
	\label{Eqn SSHMomentum}
	H_{\mathrm{SSH}}(k)= -\mathbf{d}(k) \cdot  \mathbf{\sigma},\quad k \in \left[0, 2\pi \right],
	\end{equation}
with $\mathbf{d} (k)=-\left( t_1+t_2\cos k, t_2 \sin k, 0 \right)$ and  $\mathbf{\sigma}=\left(\sigma^x, \sigma^y, \sigma^z \right)$ with $\sigma^{x, y, z}$ being the Pauli matrices.  Thus, $\mathbf{d} (k)$ defines a closed loop in the $x-y$ plane, and the Bloch vector $\mathbf{n}(k)=\mathbf{d} (k)/\left\| \mathbf{d} (k)  \right\|$ of the lower band eigenstate $\left| u_{-}(k) \right\rangle$ is exactly the radical image of $\mathbf{d} (k)$. The SSH model can thus be classified into two topologically distinct regions $\abs{g_A}<\abs{g_B}$ and  $\abs{g_A}>\abs{g_B}$, corresponding to $w=1$ and $w=0$, respectively. The variation of $g_A$ and $g_B$ can not change $w$ as long as the band gap $2\abs{\mathbf{d}(k)}$ is nonzero, that is, $\mathbf{d} (k)$ does not hit the origin in the variation. In this sense we say the topology of the Bloch band is protected by the band gap. Meanwhile, at the critical value $\abs{g_A}=\abs{g_B}$, there exists a point on $\mathbf{d} (k)$ touching the origin where the band gap is closed and $\mathbf{n}(k)$ is ill-defined. That is to say, if we want to change the topology of the Bloch band,  we need to first close the band gap at somewhere and then reopen it.

Like any realistic materials, the SSH chain does not only have a bulk part, but also has to have boundaries, i. e. two ends.The topological winding number of the Bloch band manifests itself through the existence and absence of localized zero-energy edge modes at these two ends. This embodies the important bulk-edge correspondence principle in TBT \cite{HatsugaiBEC1993PRL, HatsugaiBEC1993PRB}. We can understand this issue from the two dimerized limits $g_1=1,g_2=0$ and $g_1=0,g_2=1$ (see Fig. \ref{Fig SSH}). In both situation, ths SSH chain falls into a sequence of disconnected dimers. The difference is that each dimer is shared by the neighboring unit-cell in the topological nontrivial case, leaving two isolated, unpaired sites on the two ends and thus localized zero-energy edge modes. Moving away from this limit, the wavefunctions of the edge modes are still exponentially localized because the zero of the energy is in the bulk band gap. Therefore, the edge modes will only disappear when the band gap is closed and reopened. In this situation the topological phase transition occurs and the chain goes into the topological trivial region.

\subsection{TBT in 2D} \label{TBTin2D}
While the above 1D example is illustrative, historically the combination of the topology concept and the Bloch band theory first brought fruitful results in 2D  electronic systems, opening up the rapid growing research field topological insulators \cite{BernevigTIBook2013}.  The bulk of topological insulators behaves as an insulator while electrons can move along the edge without dissipation or back-scattering even in the presence of defect. The first example of topological insulators is the QHE  \cite{KlitzingQHE1980PRL} of 2D electrons in a uniform external magnetic field. It proved out later that the quantized Hall conductance has its root in the nontrivial topology of the bulk band structure in reciprocal space \cite{TKNN1982PRL}. A filled energy band below Fermi surface contributes a quantized portion of transverse conductance
 	\begin{equation}
	\label{Eqn TransverseConductance}
	\sigma_{h}=\frac{e^2}{2\pi\hbar}  \mathcal{C}_h,
	\end{equation}
where $h$ is the band index and $ \mathcal{C}_h$ is the corresponding topological Chern number. In particular, $ \mathcal{C}_h$ has the form
\begin{equation}
\label{Eqn ChernNumber0}
\mathcal{C}_h=\frac{1}{2 \pi} \int_{\mathrm{FBZ}} \dd^{2} k \mathcal{F}^{h}_{x y}(\mathbf{k}),
\end{equation}
where
\begin{equation}
\label{Eqn BerryCurvature}
\mathcal{F}^{h}_{x y}(\mathbf{k}) =i\left[\left\langle \frac{\partial u_h(\boldsymbol{k})}{\partial k_{x}}\left|\frac{\partial u_h(\boldsymbol{k})}{\partial k_{y}}\right\rangle-(x \leftrightarrow y)\right]\right.
\end{equation}
is the Berry curvature at $\mathbf{k}$ with $u_h(\mathbf{k})$ being the Bloch function of momentum $\mathbf{k}$ in the $h$th band. Such topological invariant is insensitive to the sample sizes, compositions, defects, and local perturbation, and can only be changed through the closing and reopening of the band gaps.   Later, it is noted that the emergence of nonzero Chern number is related  to the breaking of time-reversal symmetry, and a translational invariant Haldane model with zero net magnetic field  is proposed \cite{HaldaneQAHEPRL1988}.

\begin{figure}[tbp]
\centering
\includegraphics[width=7cm]{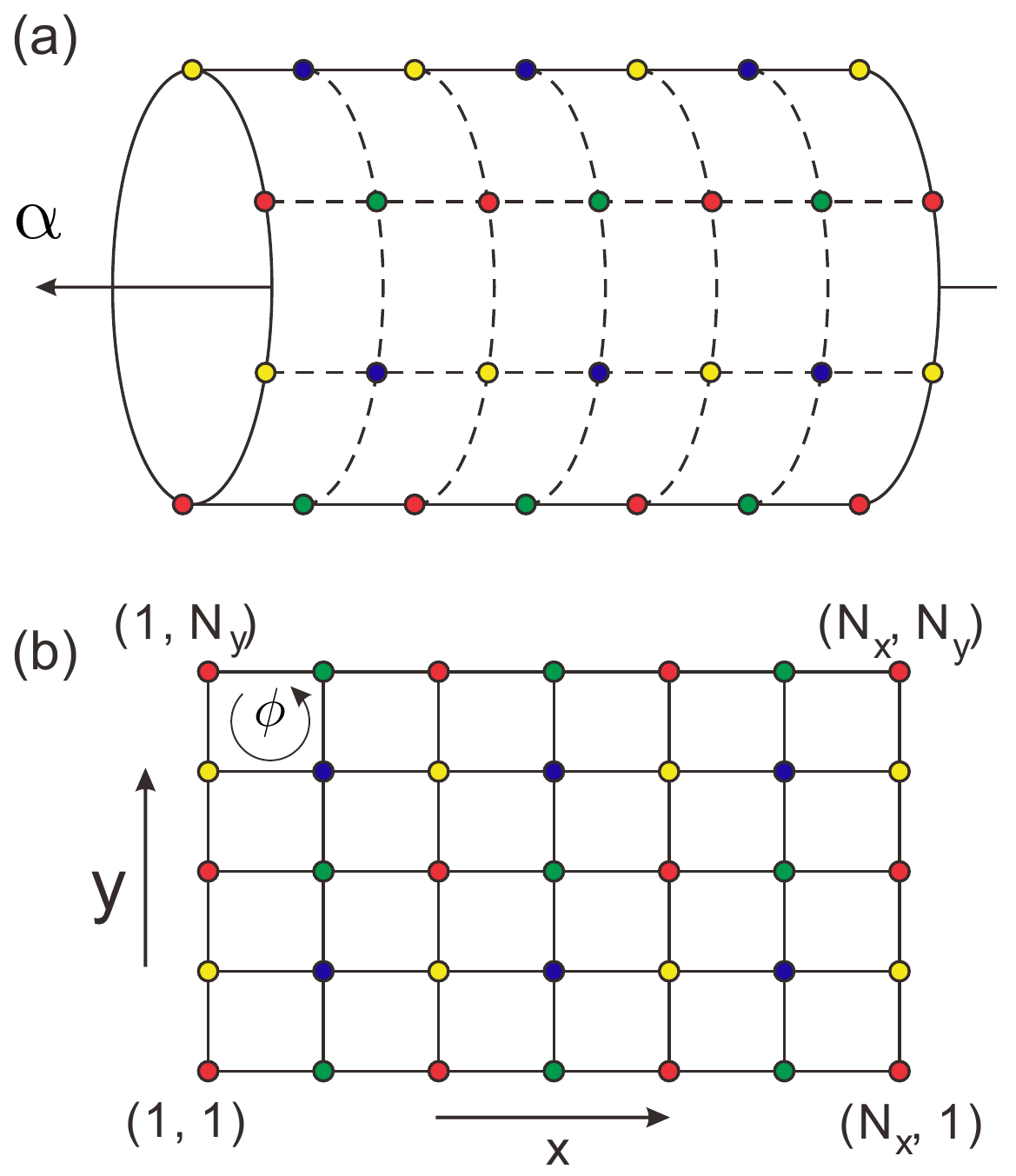}
\caption{\label{Fig Laughlinring} Sketch of  the Laughlin cylinder. The Laughlin cylinder in (a) can be regarded as a plane material wrapped along one direction (here along the y direction). Parallel to the cylinder, a magnetic flux $\alpha$ is inserted. The modulation of $\alpha$ alters the spectrum of the ESM. Through this modulation, the topological winding number of the ESMs can be extracted. Adapted from Ref. \cite{WangYPNPJQI2016}.}
\end{figure}

The 2D Chern number can also be understood in the framework of degree of map. For a two-band system with Hamiltonian $H_{\mathbf{2D}}=-\mathbf{d}(\mathbf{k})\cdot \mathbf{\sigma}$, the Berry curvature of the lowest band is calculated as \cite{QiXL2008}
\begin{align}
\mathcal{F}_{xy}(\mathbf{k})&=\frac{1}{2\left|  \mathbf{d}(\mathbf{k})  \right|^3}  \mathbf{d}(\mathbf{k}) \cdot\left[\frac{\partial \mathbf{d}(\mathbf{k})}{\partial k_{x}} \times \frac{\partial \mathbf{d}(\mathbf{k})}{\partial k_{y}}\right] \notag\\
&=\frac{1}{2}  \mathbf{n}(\mathbf{k}) \cdot\left[\frac{\partial \mathbf{n}(\mathbf{k})}{\partial k_{x}} \times \frac{\partial \mathbf{n}(\mathbf{k})}{\partial k_{y}}\right],
\end{align}
that is, $\mathcal{F}_{xy}(\mathbf{k})$ is the Jacobian of the following map
 \begin{align}
  \mathbf{k}&=\left[ k_1, k_2 \right] \notag\\
\Rightarrow \mathbf{d(k)}&=\left[ d_x(\mathbf{k}), d_y(\mathbf{k}),d_z(\mathbf{k}) \right] \notag\\
\Rightarrow \mathbf{n(k)}   &=\frac{\mathbf{d(k)}}{\left\| \mathbf{d(k)} \right\|},
 \end{align}
from the first Brillouin zone $T^2$ to $S^2$. In this sense, the integration in Eq. (\ref{Eqn ChernNumber0}) counts the times $T^2$ wraps around $S^2$ under this map. The degeneracy point $\mathbf{d}=0$  is equivalent to a fictitious monopole with topological charge $1/2$. In this sense, the $\mathbf{d}(\mathbf{k})/\left[2\left|  \mathbf{d}(\mathbf{k})  \right|^3\right]$ factor represents the field generated by the monopole, the $\frac{\partial \mathbf{d}(\mathbf{k})}{\partial k_{x}} \times \frac{\partial \mathbf{d}(\mathbf{k})}{\partial k_{y}}$ factor counts the infinitesimal  area on the two-dimensional surface $\mathbf{d(k)}$, and the quantized value of the integration of $\mathcal{F}_{xy}(\mathbf{k})$ coincides exactly with the celebrated Gauss'  law in electromagnetics.

\subsection{Topological photonics}

Here, topological photonics means simulating the topological phases on superconducting quantum circuits. However, the transfer from electronics to photonics also imposes several challenges in theory and experiments. An obvious and important issue is how to measure the topological invariants. For electronic systems, the topological invariant of the bands is measured through the quantized Hall conductance in transport experiments. This method is nevertheless not applicable in photonic systems due to the absence of the fermionic statistics. In the optics community, this problem is overcome through the realization of the gedanken adiabatic pumping process proposed by R. B. Laughlin \cite{LaughlinBEC1981PRB, HafeziWinding2014PRL, WNexperiment2016NP, WangYPNPJQI2016}. This method takes the advantage of photonic platforms that the ESMs can be manipulated and visualized in a way much better than in conventional electronic materials. We consider a 2D lattice with a uniform perpendicular magnetic field $\phi$, as shown in Fig. \ref{Fig Laughlinring}, where periodic boundary condition in the $y$-direction but open edges on the sample in the $x$-direction are placed. Due to the periodic boundary condition in the $y$-direction, a thought experiment can be devised in which the sample is wrapped in the $y$-direction into a cylinder, with $x$ being parallel to the axis of the cylinder. Therefore, the uniform magnetic flux $\phi$ becomes radial to the cylinder and ESMs emerge at the two edges. Through the cylinder and parallel to the $x$-axis, a magnetic flux $\alpha$ is inserted. The flux $\alpha$ can shift the momentum of the ESMs.  When one flux quanta is threaded through the Laughlin cylinder exactly, the ESM spectrum return to its original form with an integer number of ESMs being transferred. This integer is the winding number of the ESMs, which is equivalent to the Chern number of the bulk bands.

\section{Basics of  superconducting quantum circuits}
\label{Sec Para}
In this section we briefly introduce the basic concepts of SQC. For the purpose of our review, we put special emphasis on the 2D superconducting TLR and superconducting transmon qubit and the parametric coupling mechanism based on these SQC elements.

\subsection{Superconducting TLR}
A TLR consists of a coplanar waveguide formed by a center conductor separated on both sides from a ground plane.  This planar structure can be regarded as a 2D realization of  conventional coaxial cable 
\cite{YaleCQEDPRA2004}. The characteristic electromagnetic parameter of the coplanar waveguide are its characteristic impedance $Z_{r}=\sqrt{l_{0} / c_{0}}$ and the speed of light in the waveguide $v_{0}=1 / \sqrt{l_{0} c_{0}}$ , with $c_0$ and $l_0$ being the capacitance to ground and the inductance $l_0$ per unit length. Typical values of these parameters are $Z_{r} \sim 50 \Omega$ and $v_0 \sim 1.3 \times \mathrm{10^8 m/s}$ \cite{WallraffNature2004}. In the continuum limit, the Hamiltonian density in the bulk of the wave guide has the form
\begin{equation}
\mathcal{H}_{\mathrm{TLR}}=\frac{1}{2 c_{0}} Q(x,t)^{2}+\frac{1}{2 l_{0}}\left[\partial_{x} \Phi(x,t)\right]^{2},
\label{Eqn TLRHamiltonian}
\end{equation}
where $Q(x,t)$ is the charge density distribution function on the waveguide and the generalized flux $\Phi(x, t)=\int_{-\infty}^{t} d t^{\prime} V\left(x, t^{\prime}\right)$ is the canonical conjugate of $Q(x,t)$ with $V(x, t)$ being the voltage to ground on the center conductor. By using  Hamilton's equations, the wave equation describing the surface plasma propagation along the transmission line,
\begin{equation}
v_{0}^{2} \frac{\partial^{2} \Phi(x, t)}{\partial x^{2}}-\frac{\partial^{2} \Phi(x, t)}{\partial t^{2}}=0,
\end{equation}
can be derived.

A resonator mode, or a cavity mode, is further establish from Eq. \eqref{Eqn TLRHamiltonian} by
imposing boundary conditions at the two end points of the waveguide, which can be
achieved by microfabricating a gap in the center conductor. A TLR has a fundamental frequency $\omega_0=\pi v_0/d$ and harmonics at $\omega_m=m\omega_0$ with $d$ being the length of the waveguide. Therefore, we get a single-mode $\lambda/2$ TLR resonator if we focus only on the lowest mode of the coplanar waveguide. The similar derivation can also be applied to the shorted boundary condition considered in the following section where the two ends are grounded. It can be noticed that such coplanar waveguide geometry is flexible and a large
range of eigenfrequency $\omega_0$ can be achieved. Typical SQC experiments use TLR resonators with $\omega_0 \in [5, 15] \mathrm{GHz}$, which is much larger than the thermal temperature in the dilute refrigerator  ($\sim 20 \mathrm{mK}$) such that thermal excitation is suppressed, and much smaller than the superconducting energy gap of aluminum or niobium such that the unwanted quasiparticle excitation can be safely neglected.
In addition, in experiments it is always advantageous to maximize the internal quality factor of the TLR mode. With current level of technology, the loss of the TLR mode due to its coupling to uncontrolled degrees of freedom in the environment can now be effectively suppressed, and internal Q-factor in the range $[10^5,10^6]$ can now be routinely achieved \cite{LiuYXReview2017PR, CQED2021}.

\subsection{Superconducting transmon qubit}
\begin{figure}[tbp]
\centering
\includegraphics[width=7cm]{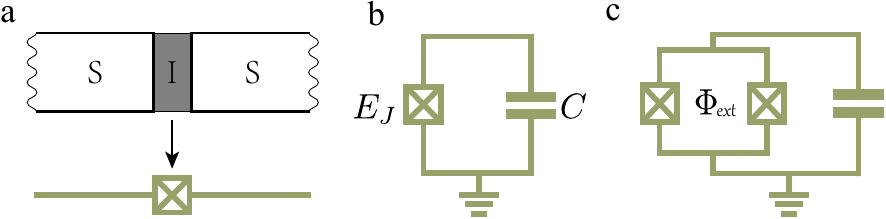}
\caption{\label{Fig Transmon} Schematic plot of (a) a Josephson junction consisting of a weak link between two spuerconducting islands, (b) a transmon qubit formed by a Josephson junction with maximal Josephson coupling energy $E_J$ shunted by a large capacitance, and (c) a frequency-tunable transmon qubit, with the single junction replaced by a SQUID penetrated by external magnetic flux $\Phi_{ext}$. }
\end{figure}

The described 2D TLR is a linear circuit. However, nonlinearity is essential to establish two-level qubits in which quantum information is encoded and manipulated. Meanwhile, superconductivity allows the introduction of non-dissipative nonlinearity in mesoscopic electrical circuits as Josephson junction can be regarded as a nonlinear inductance. For two superconducting metals separated
by a thin insulating barrier, a dissipationless supercurrent could flow between them, described by the relation \cite{TinkhamBook2004}
\begin{equation}
I=I_{c} \sin \varphi,
\end{equation}
with $I_c$ being the critical current of the Josephson junction, and $\varphi$ is the phase difference between the superconducting metals on the two sides of the junction (see Fig. \ref{Fig Transmon}(a)). Moreover, the time dependence of $\varphi$ is related to the voltage across the junction as
\begin{equation}
\frac{d \varphi}{d t}=\frac{2 \pi}{\Phi_{0}} V,
\end{equation}
with $\Phi_0=\pi\hbar/2$ being the flux quantum. Here $\varphi$ can be associated with the previously introduced generalized flux as
$\varphi(t)=2 \pi \Phi(t) / \Phi_{0}=2 \pi \int d t^{\prime} V\left(t^{\prime}\right) / \Phi_{0}$.  Therefore, we can regard the Josephson junction as a nonlinear inductance
\begin{equation}
L_{J}(\Phi)=\left(\frac{\partial I}{\partial \Phi}\right)^{-1}=\frac{\Phi_{0}}{2 \pi I_{c}} \frac{1}{\cos \left(2 \pi \Phi / \Phi_{0}\right)}.
\end{equation}
With the nonlinear Josephson junction inductance, we can build a nonlinear LC resonator by replacing the inductance of a linear LC oscillator with a Josephson junction. In this situation, the energy levels of the nonlinear LC resonator are no longer equidistant, and we can get an effective two level qubit if we restrict our attention to the lowest two energy levels.

More explicitly, we consider the circuit diagram of a transmon qubit depicted in Fig. \ref{Fig Transmon}(b), which consists of a Josephson junction shunted by a large capacitance $C$. Since the energy stored in the Josephson junction can be written as
 \begin{equation}
E= \int IVd t=-E_{J} \cos \left(\frac{2 \pi}{\Phi_{0}} \Phi\right)
\end{equation}
 with $E_{J}=\Phi_{0} I_{c} / 2 \pi$ being the Josephson energy of the junction, the quantized Hamiltonian of the capacitively
shunted Josephson junction therefore reads
\begin{eqnarray}
{H}_\mathrm{Transmon}
&=& \frac{Q^2}{2 C}-E_{J} \cos \left(\frac{2 \pi}{\Phi_{0}}  {\Phi}\right)\notag\\
&=& 4 E_{C}{n}^{2}-E_{J} \cos{\varphi},
\end{eqnarray}
where $C$ denotes the total capacitance around the transmon qubit, $E_{C}=e^{2} / 2 C$ is the charging energy, and $n=Q/2e$ is the canonical momentum of $\varphi$ describing the excess number of Cooper pairs on the metallic island. The spectrum of ${H}_\mathrm{Transmon}$ depends on the ratio $E_{J} / E_{C}$. It has been shown that, in the range $E_{J} / E_{C} \in [20,80]$, the energy levels of ${H}_\mathrm{Transmon}$ are insensitive to the perturbation from the low-frequency $1/f$ background charge noise \cite{KochTransmonPRA2007}. In this situation ${H}_\mathrm{Transmon}$ describes a virtual particle with generalized coordinate $\varphi$, generalized momentum $n$, and effective mass $(8E_C)^{-1}$, moving in an anharmonic potential $-E_{J} \cos{\varphi}$. In turn, ${H}_\mathrm{Transmon}$ can have an anharmonic oscillator form
\begin{eqnarray}
{H}_\mathrm{Transmon}  \approx \hbar \omega_{q} {b}^{\dagger} {b}-\frac{E_{C}}{2} {b}^{\dagger} {b}^{\dagger} {b} {b},
\end{eqnarray}
where $b$($b^\dagger$) is the annihilation(creation) operator of the anharmonic oscillator, and $\hbar \omega_{q}=E_1-E_0=\sqrt{8 E_{C} E_{J}}-E_{C}$ denotes the energy spacing of the lowest two levels ($E_n$ being the eigenenery of the Fock state $\ket{n}$). The anharmonicity of the resonator is described by the level difference $-E_C=(E_2-E_1)-(E_1-E_1)$. Such anharmonicity factor sets the upper speed limit of manipulating the lowest two levels.

In addition, while the critical current $I_c$ depends on the junction size and the material parameters, we can have tunable effective critical current by replacing the single junction by a superconducting quantum interference device (SQUID) loop formed by two identical junctions and penetrated by an external flux $\Phi_{ext}$, as shown in Fig. \ref{Fig Transmon}(c). In this situation the SQUID can be regarded as a Josephson junction with tunable critical current $I_c(\Phi_{ext})=2I_c\cos(\pi\Phi_{ext}/\Phi_0)$ with $\Phi_0=\pi\hbar/2$ being the flux quanta. Such tunability origins from the Aharonov-Bohm interference of the supercurrent flowing across the two junctions. Therefore, replacing the single junction by a SQUID loop yields directly a transmon qubit with flux-tunable eigenfrequency. Moreover, as the flux in the SQUID loop can be tuned with a very high frequency in practice, this additional control knob results in  the parametric driving mechanism which we will discuss in the subsequent sections.

\subsection{Static  couplings}
The described TLR resonators and transmon qubits can be connected to form a scaled lattice, and their individual modes can be used to play the role of localized Wannier modes of the lattice. To perform meaningful quantum simulation, one has to induce the coupling between the connected SQC elements. An intuitive idea is to connect them directly by coupling capacitances or inductances. Since the voltages and currents on the SQC elements can be represented by the mode creation and annihilation operators as forms like $V=V_{ZPF}(a+a^\dagger)$ and $I=I_{ZPF}(ia-ia^\dagger)$ (ZPF for zero-point fluctuation), this direct coupling mechanism results in exchange type coupling $\sim C_{coupling}V_1V_2= g(a^\dagger b+ b^\dagger a)$ between SQC elements. However,  it should be notice that the resulting coupling constant can not be tuned either in amplitude or in phase. Actually, this kind of coupling results in only static and completely real coupling constants. This is particularly not desirable because complex coupling constants between SQC elements is needed to break the time reversal symmetry and obtain the consequent nontrivial topology. Also,  a method exploiting the double electromagnetically induced transparency scheme \cite{BoseFQHEPRL2008, YangWLGaugeFieldPRA2012} to realize effective magnetic fields for polaritons in a 2D cavity lattice has been proposed. However, the using of multi-level artificial atoms there are still challenging with current experimental technologies.

\section{Parametric couplings}
As the static coupling is not tunable, we now introduce an alternative method based on the mechanism of parametric frequency conversion (PFC), which can induce coupling between lattice sites with controllable strengths and phases, and discuss a variety of its consequent topological effects. This method is inspired by the laser assisted tunneling technique invented in atomic lattices \cite{ZollerLaser2003}. With appropriate modulating frequency imposed on either the couplers between SQC elements or the on-site eigenfrequencies of the SQC elements, effective parametric coupling between SQC elements can be induced, where a hopping phase accompanies during the hopping process, and from the hopping phases, an effective magnetic field on the microwave photons can be induced.

\subsection{Parametric couplings among TLRs}

We first consider how to establish the tunable photonic hopping between the lattice sites by using the PFC method \cite{NISTParametricConversionNP2011, NISTCoherentStateAPL2015, NISTHongOuMandelPRL2012, GrossPFCEPJQT2016}. The essential physics can be illustrated intuitively through a toy model of two coupled cavities, assuming $\hbar\equiv1$ hereafter, with a Hamiltonian of
\begin{equation}
{H}_{\mathrm{TC}}={H}_{0}+{H}^{12}_{\mathrm{ac}}(t),\label{equ:model2}
\end{equation}
with
\begin{align}
&  {H}_{0}=\sum_{i=1}^{2}\omega_{i} a_{i}^{\dag}a_{i},\\
&  {H}^{12}_{\mathrm{ac}}(t)=g_{12}(t)[a_{1}^{\dag}+a_{1}][a_{2}^{\dag}+a_{2}],
\end{align}
where $a_{i}/a_{i}^{\dag}$ and $\omega_i$ are the annihilation/creation operators and the eigenfrequencies of the two cavities, respectively. In addition, the form of ${H}^{12}_{\mathrm{ac}}(t)$ roots from the inductive current-current coupling of two TLRs \cite{GrossPFC2013PRB,GrossPFCEPJQT2016}, and the time-dependent coupling constant $g_{12}(t)$ corresponds to the ac modulation of the coupling SQUID between the TLRs.

Our aim is to implement a controllable inter-TLR photon hopping process. As the two cavities are far off-resonant, the required photon hopping can hardly be achieved  if $g_{12}(t)$ is static. Meanwhile, the effective $1\leftrightarrow2$ photon hopping can be implemented by modulating $g_{12}(t)$ dynamically as
\begin{eqnarray}
g_{12}(t)&=& 2J\cos[(\omega_{1}-\omega_{2})t-\theta_{12}]\notag\\
&=& J\left\{e^{i[(\omega_{1}-\omega_{2})t-\theta_{12}]} + \mathrm{h.c.} \right\}.
\label{equ:drivingsimple}
\end{eqnarray}
Explicitly, under the assumption of  $(\omega_{2}+\omega_{1}) \gg |\omega_{2}-\omega_{1}| \gg J$, the effective Hamiltonian in the interaction picture with respect to $U_0=\exp\{-\text{i} {H}_{0} t\}$ takes the form
\begin{eqnarray} \label{equ:model}
{H}_{\mathrm{eff}}^{12}&=& U_0^\dagger {H}^{12}_{\mathrm{ac}}(t) U_0\notag\\
&=& g_{12}(t)[a_{1}^{\dag}e^{i\omega_1 t}+a_{1}e^{-i\omega_1 t}][a_{2}^{\dag} e^{i\omega_2 t} +a_{2} e^{-i\omega_2 t}]\notag\\
&=& g_{12}(t)[a_{1}^{\dag}a_{2}^{\dag}  e^{i(\omega_1+\omega_2) t}+a_{1}^{\dag}a_{2}   e^{i(\omega_1-\omega_2) t} + \mathrm{h.c.}]\notag\\
&\approx & g_{12}(t)[a_{1}^{\dag}a_{2}   e^{i(\omega_1-\omega_2) t} + \mathrm{h.c.}]\notag\\
&=& J\left\{e^{i[(\omega_{1}-\omega_{2})t-\theta_{12}]} + \mathrm{h.c.} \right\} [a_{1}^{\dag}a_{2}   e^{i(\omega_1-\omega_2) t} + \mathrm{h.c.}]
\notag\\
&=&  J  a_{1}^{\dag}a_{2} \{ e^{i\theta_{12}} +    e^{i[2(\omega_1-\omega_2) t-\theta_{12}]} + \mathrm{h.c.}\}\notag\\
&\approx & J  a_{1}^{\dag}a_{2}  e^{i\theta_{12}} + \mathrm{h.c.},
\end{eqnarray}
with $J$ and $\theta_{12}$ being the effective   hopping rate and the hopping phase, respectively. Physically speaking, we can imagine that there is a photon initially placed in the 1st cavity. As $g_{12}(t)$ carries energy quanta filling the gap between the two cavity modes, the photon can absorb the needed energy from the oscillating coupling strength $g_{12}(t)$,  convert its frequency to $\omega_{2}$, and finally jump into the 2nd cavity. During this hopping, the initial phase of $g_{12}(t)$ is adopted by the photon. This hopping process can be further described in a rigorous way by using the rotating wave approximation: in the rotating frame of ${H}_{0}$, ${H}^{12}_{\mathrm{ac}}(t)$ reduces to the form in Eq. (\ref{equ:model}), with the fast oscillating  $a_1^\dagger a_2^\dagger+\mathrm{h.c.}$ term neglected. The essential advantage of the described PFC method is that both the effective hopping strength $J$ and the hopping phase $\theta_{12}$ can be controlled on-demand by the modulating pulse $g_{12}(t)$. In particular, the implementation of the hopping phase $\theta_{12}$ is important as we are using this method to synthesize artificial gauge fields.

From the above derivation, it becomes clear that the essential of realizing this parametric coupling formalism is to realize the tunable coupling constant $g_{12}(t)$. Here we remember that a SQUID can be regarded as a Josephson junction with tunable critical current controlled by the external bias flux. If we can couple the currents from two SQC elements by a SQUID, and if the SQUID is working in the linear region, we can have a tunable coupling between these two SQC elements because in this situation the coupling SQUID can be regarded as a tunable inductance. In the next section we come back to the explicit circuit realization of this idea.

\subsection{Parametric tunable coupling among qubits}

In most quantum information processing tasks, tunable coupling between qubits is  of particular importance. However, this tunability is usually achieved  at the cost of introducing additional decoherence or circuit complexity \cite{devicecouple1, devicecouple2, RoushanChiral2017NP,devicecouple4, devicecouple5, devicecouple6, devicecouple7}. Alternatively, parametric modulation of qubits' frequencies can be used to realize this tunability \cite{paracouple1, paracouple2, paracouple4} without  coupling devices, and thus simplifies the circuit.

\begin{figure}\begin{center}
\includegraphics[width=0.45\textwidth]{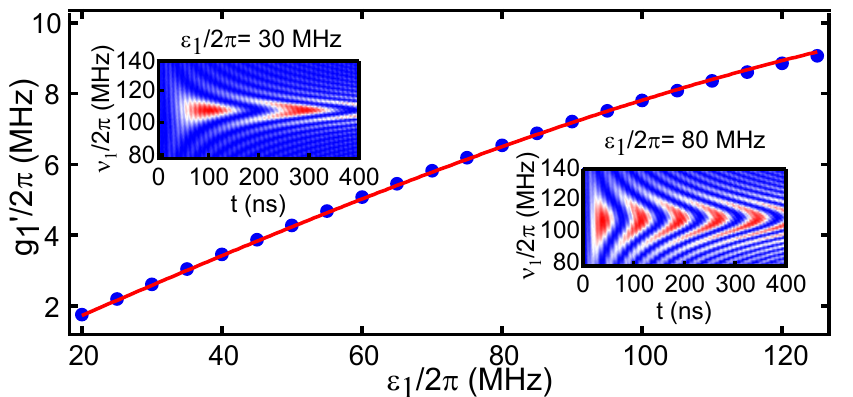}
\end{center}
\caption{ Experimental parametric tunable coupling $g_1'$ between $Q_0$ and $Q_1$ as a function of $\varepsilon_1$, where $Q_0$ is operated at a fixed frequency $\omega_{\mathrm{0}}$ and initially prepared in its first excited state $\ket{e}$, while $Q_1$ is flux biased in a time-dependent way so that its frequency is oscillating in the form as in Eq. (\ref{Eq2}) and initially prepared in $\ket{g}$. Coherent excitation oscillation between $Q_0$ and $Q_1$ as a function of $\nu_1$ and time $t$ at fixed $\varepsilon_1$ produces a chevron pattern, from which the corresponding $g_1'$ (dots) can be extracted, i.e., the frequency of the full-scale oscillation ($\nu_1=\Delta_1$) pattern is the corresponding $g_1'$. Two typical chevron patterns are presented in the inset, with blue and red corresponding to $\ket{g}$ and $\ket{e}$  of $Q_1$, respectively.   The red line is fitted using Eq.~(\ref{paracoupleg}) with $g_1/2\pi   \approx 19$~MHz. Adapted from Ref. \cite{paracouple2}. }
\label{figparacoupleg}
\end{figure}

For the parametrically tunable coupling, we here first review the 1D chain case \cite{paracouple2}. This tunability  method  can also be directly applicable to the time-dependent tuning \cite{paracouple4} and 2D \cite{twoleg} cases. The system Hamiltonian for \emph{N} 1D capacitively coupled transmon qubits  \cite{KochTransmonPRA2007, transmon2}  is
\begin{eqnarray}\label{Eq1}
H=\sum^{N-1}_{j=0}\frac {\omega_j} {2}\sigma^z_j+\sum^{N-1}_{j=1}g_j \sigma^x_{j-1} \sigma^x_{j},
\end{eqnarray}
where  $\sigma^{z, x}_j$ are the Pauli operators on the $j$th qubit $Q_j$ with transition frequency $\omega_j$, and $g_j$ is the static coupling strength between qubits $Q_{j-1}$ and $Q_{j}$, which is fixed. The coupling strength can be fully tuned via  modulating the qubits $Q_j$ with $1\leq j\leq{N-1}$ so that the frequencies of the qubits are oscillating as
\begin{eqnarray}\label{Eq2}
\omega_j=\omega_{\mathrm{o}j}+\varepsilon_j \sin(\nu_jt+\varphi_j),
\end{eqnarray}
where $\omega_{\mathrm{o}j}$ is the fixed  operating qubit-frequency, $\varepsilon_j$, $\nu_j$, and $\varphi_j$ are the modulation amplitude, frequency, and phase, respectively. When $\Delta_j=\omega_{\mathrm{o}j}-\omega_{\mathrm{o}(j-1)}$ equals to $\nu_j$ ($-\nu_j$) for odd (even) $j$, in the interaction picture and ignoring the higher-order oscillating terms, we get a chain of qubits and the interaction among then is the nearest-neighbor resonant $XY$ coupling i.e.,
\begin{eqnarray}\label{eq:H}
H_I =\sum^{N-1}_{j=1} g^\prime_j \sigma^+_{j-1}\sigma^-_j+\text{H.c.},
\end{eqnarray}
where $\sigma^\pm=(\sigma^x\pm\sigma^y)/2$ and the effective tunable coupling strengths are
\begin{eqnarray}\label{paracoupleg}
g^\prime_j= g_j J_1(\eta_j)\times
\left\{
  \begin{array}{ll}
   e^{i(\varphi_1+\pi/2)}, & \hbox{$j=1$;} \\
    J_0(\eta_{j-1}) e^{-i(\varphi_j-\pi/2)}, & \hbox{$j$ is even;}\\
   J_0(\eta_{j-1}) e^{i(\varphi_j+\pi/2)}, & \hbox{$j$ is odd and $\neq$ 1},
      \end{array}
\right.
\end{eqnarray}
with $\eta_j=\varepsilon_j/\nu_j$, $J_m(\eta_j)$ being the $m$th Bessel function of the first kind. In this way, the  tunability of $g^\prime_j$ is achieved by changing the amplitude $\varepsilon_j$ of the modulation to tune $\eta_j$ , as experimental verified in Fig. \ref{figparacoupleg} for a two-qubit case.

For tuning the coupling strength via an auxiliary device,  the current wisdom is to capacitively   coupled to two weakly capacitively coupled computational superconducting qubits  via a tunable coupler \cite{devicecouple4}, which is actually a superconducting qubit, and thus this architecture is easy to scale up. In this scenario,  the qubit-qubit coupling will attribute to two paths, one is the original fixed weak capacitively coupling, the other is induced from the stronger large detuned qubit-coupler interaction, which can be adjustable  by tuning the frequency of the coupler. In this way, the total qubit-qubit coupling is still in the form as that of Eq. (\ref{eq:H}), and it can also be fully tuned in a continuous way, even from positive to negative values \cite{devicecouple5}. Alternatively, the cross-Kerr ZZ-interaction can be obtained for two qubits with large qubit-frequency difference, and, in this case,  controlled-phase gates can be implemented \cite{devicecouple6, devicecouple7}. As the total interaction is  adjusted from large detuned qubit-coupler interaction, the unwanted qubit interactions can also be completely turned off, and thus high-fidelity quantum operations can be possible.

\section{TBT with TLR}

In this section, we present the idea of  using PFC between TLRs to establish SQC lattice and observe the consequent topological effects.

\subsection{TBT in quasi 1D}

\begin{figure}[tb]
\begin{center}
\includegraphics[width=0.45\textwidth]{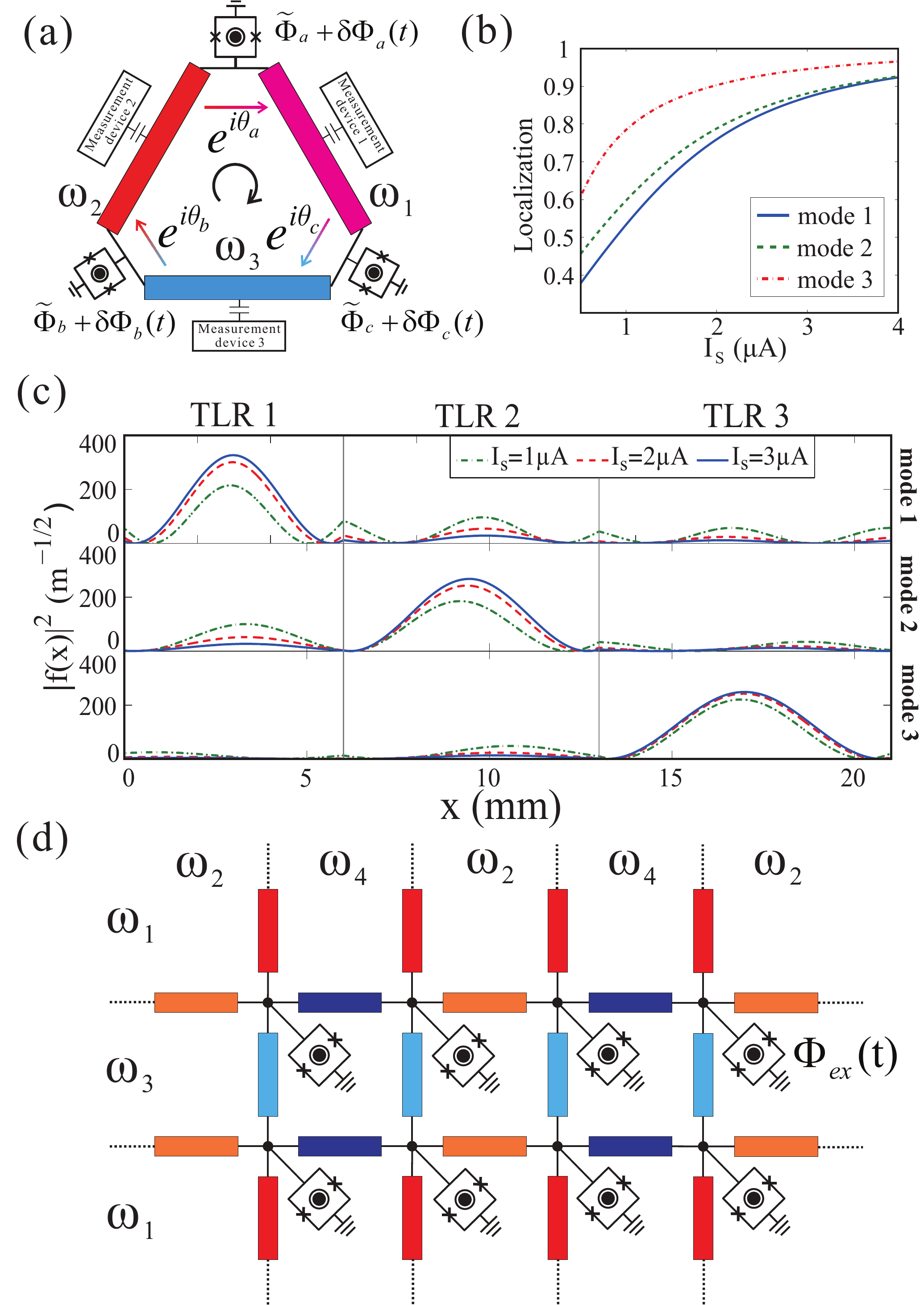}
\end{center}
\caption{\label{Fig Necklace}  (a) Schematic plot of the three-TLR necklace coupled by grounding SQUIDs.  The loop of each SQUID is penetrated by a static bias $\widetilde{\Phi}$ and an ac modulating pulse $\delta\Phi$. (b), (c) The localization property of the eigenmodes. In (c),  the normalized mode functions of the lowest eigenmodes of the necklace versus the critical current $I_S$ of the grounding SQUIDs are depicted.  The three panels of (c) describe the mode functions of the eigenmodes 1, 2, and 3 from top to bottom. In (b), we quantify the localization property of the $m$th eigenmodes by the energy stored in the $m$th TLR versus the total eigenenergy of the $m$th mode. Adapted from Ref. \cite{WangYPChiral2015}.}
\end{figure}

We then illustrate explicitly the implementation of the described PFC method in SQC system. In particular, we consider how to synthesize artificial gauge field for the microwave photons with this method, and  a variety of the consequent novel topological effects. Our first work in this direction is to investigate the chiral photon flow effect  in a TLR necklace \cite{WangYPChiral2015}, which has been demonstrated experimentally by using three coupler coupled superconducting transmon qubits \cite{RoushanChiral2017NP}.  As shown in Fig. \ref{Fig Necklace}(a), the proposed circuit consists of three TLRs coupled by three grounding SQUIDs with very large effective Josephson coupling energies and can be described by
\begin{equation}
\mathcal{H}_{0}=\sum_{m=1}^{3}\omega_{m} a_{m}^{\dag}a_{m},
\label{equ:Hamiltonianchiral}
\end{equation}
where $a_m$ are the annihilation operators of the $\lambda/2$ modes of the three TLRs and $\omega_m$ are the corresponding eigenfrequencies. Here the parameters are specified as $(\omega_1,\omega_2,\omega_3)=(\omega_{\mathrm {0}},\omega_{\mathrm{0}}-\Delta, \omega_{\mathrm{0}}+2\Delta)$ with $\omega_{\mathrm{0}}/2\pi \in \left[10,15\right]\,\mathrm{GHz}$ and $\Delta/2\pi \in \left[1,2\right]\,\mathrm{GHz}$. Such configuration is for the following application of the PFC scheme and can be achieved in experiment through the length selection of the TLRs \cite{NISTParametricConversionNP2011, NISTHongOuMandelPRL2012, NISTParametricCouplingPRL2014, NISTCoherentStateAPL2015}.
The mode structure of the lattice can be explained in an intuitive way: from the point of view of a particular TLR, its grounding SQUIDs act as shortcuts of its neighbors and itself, as most of the currents from the neighboring TLRs will flow directly to the ground through their common grounding SQUIDs without crossing each other. The separated and localized modes for the TLR lattice can then be approximated by the individual $\lambda /2$ modes of the TLRs, as the small inductances of the grounding SQUIDs impose the grounding nodes at the edges of the TLRs.

\begin{figure}[tb]
\begin{center}
\includegraphics[width=0.48\textwidth]{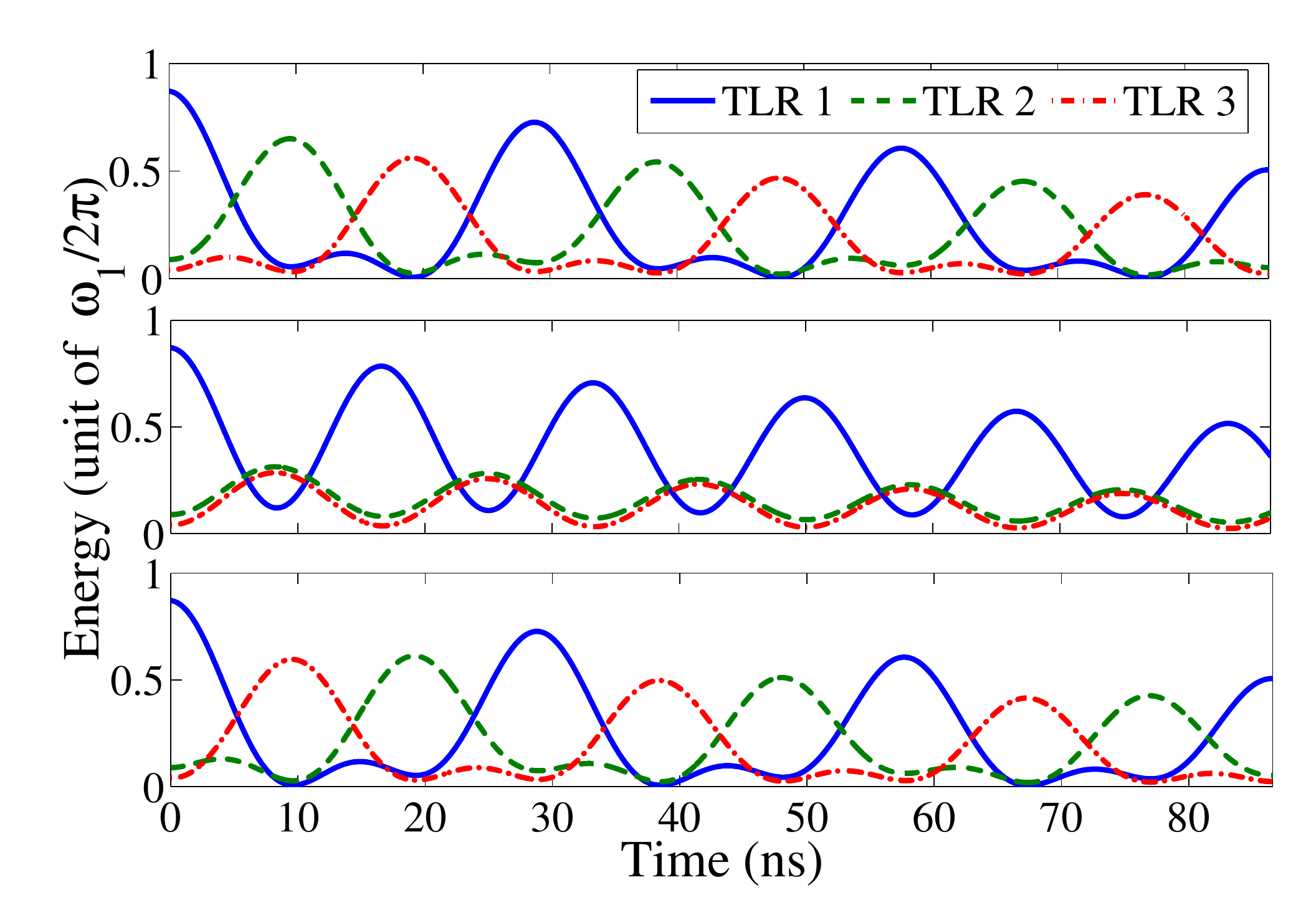}
\end{center}
\caption{Chiral photon flow in the TLR necklace. We have considered three situations $\theta_{\Sigma}=\pi/2$, $\pi$, and
$3\pi/2$, and depict the corresponding results in the three panels from top to bottom. In addition, we set the homogeneous coupling strength $g/2\pi=20$ \textrm{MHz} and decay rate $\kappa/2\pi=250$~kHz. Adapted from Ref. \cite{WangYPChiral2015}.}
\label{Fig ChiralFlow}
\end{figure}

We turn to the issue of establishing the inter-TLR coupling through the described PFC method. First we take the $1\leftrightarrow 2$ coupling as an example. Physically speaking, the common grounding SQUID can be regarded as a tunable mutual inductance between these two TLRs, because currents from these two TLRs flow to the ground through the same grounding SQUID. Therefore, we add an ac flux driving $\delta
\Phi _{12}(t)=\Delta \Phi _{12}\cos [(\omega _{1}-\omega _{2})t-\theta _{12}]$
to modulate the effective inductance of the $1\leftrightarrow 2$ SQUID, such that  the parametric coupling Hamiltonian
\begin{equation}
\mathcal{H}_\mathrm{int,12}=2J \cos [(\omega _{1}-\omega _{2})t-\theta
_{12}]a_{1}^{\dag }a_{2}+\mathrm{h.c.},  \label{equ:Hamiltonianchiral2}
\end{equation}
can be induced with $J$ being the coupling strength proportional to $\Delta \Phi _{12}$.  In the rotating frame of $\mathcal{H}_{0}$ in Eq. (\ref{equ:Hamiltonianchiral}), an effective $1\leftrightarrow 2$ hopping process described by
\begin{equation}
\mathcal{H}_\mathrm{int,12}=J e^{i\theta_{12}}a_{1}^{\dag }a_{2}+\mathrm{h.c.},
\label{equ:chiralHamiltonian}
\end{equation}
can then be obtained, which exactly reproduced Eq. (\ref{equ:model}). Moreover, we can add similar pumping pulses on the other two SQUIDs and finally get the effective
Hamiltonian%
\begin{equation}
\mathcal{H}_{I}=J\left[e^{i\theta _{12}}a_{1}^{\dag }a_{2}+e^{i\theta
_{23}}a_{2}^{\dag }a_{3}+e^{i\theta _{31}}a_{3}^{\dag }a_{1}\right]+\mathrm{H.c.}.
\label{equ:chiralHamiltonian2}
\end{equation}
Here the hopping strength $J$ and the three hopping phases $\theta_{12}$, $\theta_{23}$, and $\theta_{31}$ can be modulated by the amplitudes and the initial phases of the modulating pulses, respectively. In this step, the vector potential $\mathbf{A(x)}$ can manifest its presence through the Peierls substitution
\begin{equation}
\theta_{ij}=\int_{j}^{i} \mathbf{A(x)}\cdot\mathrm{d}\mathbf{x},
\end{equation}
and the loop summation of the hopping phases in turn has the physical meaning of the synthetic magnetic flux:
\begin{equation}
\theta_{\Sigma}=\oint \mathbf{A(x)}\cdot\mathrm{d}\mathbf{x}=\iint \mathbf{B(x)}\cdot\mathrm{d}\mathbf{S}.
\end{equation}

The synthetic magnetic field leads to the chiral photon flow in this necklace, which is the photonic counterpart of the Lorentz circulation of electrons in an external magnetic field. We assume initially a photon is populated in the 1st mode and numerically simulate its subsequent time evolution by using the master equation approach. Results related with the three situations $\theta_{\Sigma}=\pi/2,\pi$ and $3\pi/2$ are plotted in Fig. \ref{Fig ChiralFlow}. From the first panel corresponding to $\theta_\Sigma=\pi/2$, we can observe the clear temporal phase delay of  the energy population in the three TLRs. Such pattern  implies that the photon flow on the lattice is unidirectional, first from TLR $1$ to TLR $2$ and then from TLR $2$ to TLR $3$. This chiral character is a consequence of the time-reversal symmetry breaking in this necklace.  Here we should emphasize that the chiral character of the photon flow survives although we have chosen the cavity decay rate $\kappa$ much stronger than reported experimental data during this numerical simulation, Similarly, in the third panel which corresponds to the opposite magnetic field $\theta_{\Sigma}=3\pi/2$,  the chiral photon flow is in the opposite direction. Meanwhile, in the second panel corresponding to the trivial case $\theta_\Sigma=\pi$, the energy transfer exhibit symmetric patterns.

Meanwhile,  we can notice that our discussions up to now are mainly restricted to the synthesization of and the effect induced by  Abelian background gauge  field \cite{KogutNAGRMP1983}. On the other hand, non-Abelian gauge field, in particular the spin-orbital coupling (SOC) mechanism \cite{SpielmanSOC2013Nature}, has already played important roles in the physics of topological quantum matters \cite{BernevigTIBook2013}. In recent year, synthetic SOC has been experimentally implemented in artificial systems including ultra-cold atoms \cite{PanJWSOCScience2016}, exciton-polariton microcavities \cite{AmoSOCPRX2015}, and coupled pendula chains \cite{SalernoSOC2017NJP}. Therefore, curious questions arise that whether the proposed PFC method can be extended to synthesize non-Abelian gauge in the SQC architecture, and whether the synthetic non-Abelian gauge field would bring any new physics that can be observed in the near future.

\begin{figure}[tb]
\begin{center}
\includegraphics[width=0.45\textwidth]{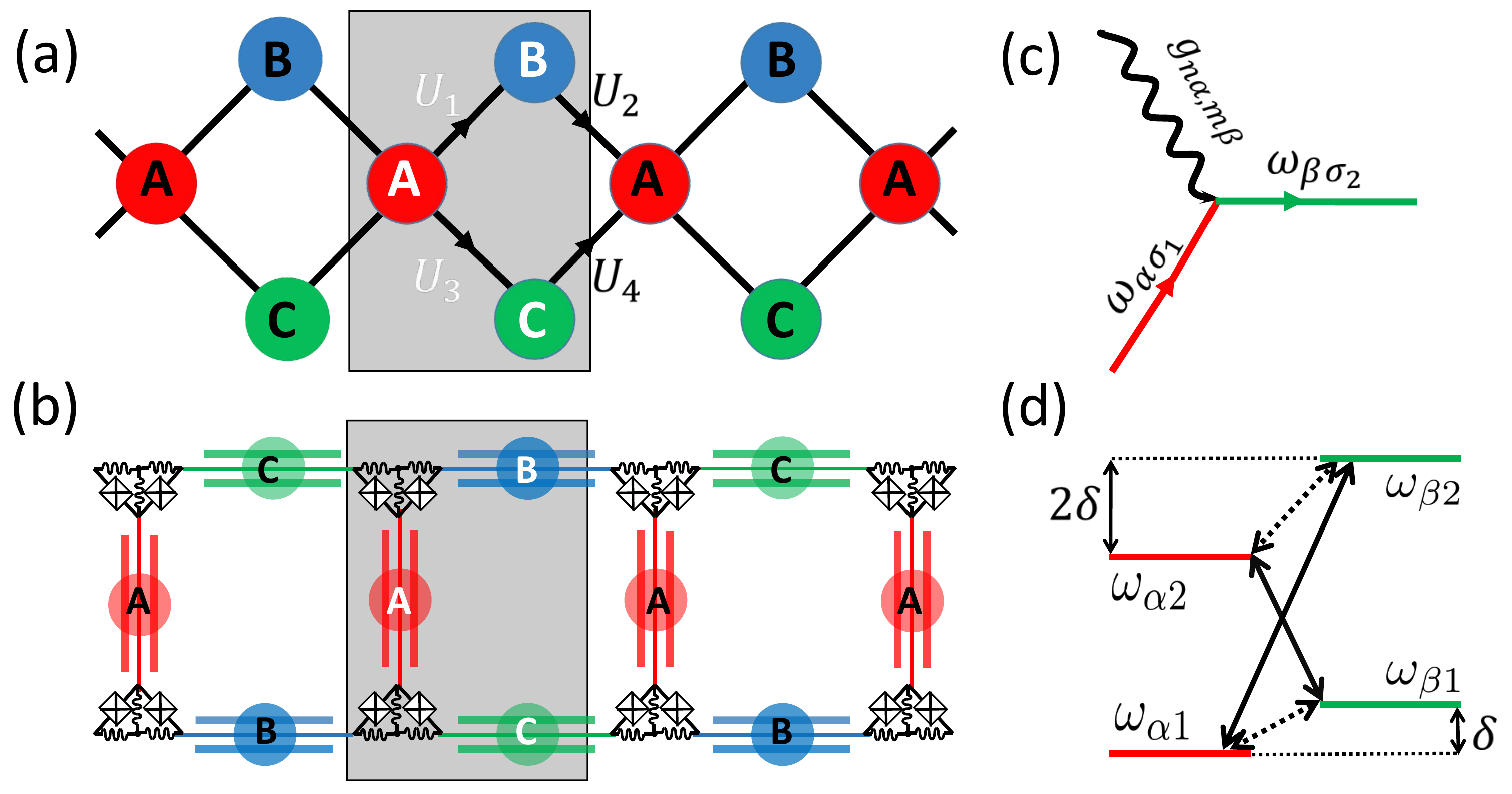}
\end{center}
\caption{ Implementation of a quasi 1D  AB caging  model. (a) Illustration of the model in a periodic rhombic lattice.  (b) The extension of (a) to a two component non-Abelian  case and its implementation with TLR lattice, where the TLRs with three different lengths are connected at the common ends by the grounded coupling SQUIDs. (c) Illustration of the effective $[n,\alpha,\sigma_1]\Leftrightarrow[m,\beta,\sigma_2]$ coupling induced by  the PFC via the ac modulation of the coupling SQUID. (d) Spectrum and coupling of two neighboring TLR sites, with $\delta=\omega_{\beta 1}-\omega_{\alpha 1}$ being the frequency difference of the neighboring $[n,\alpha]$ and $[m,\beta]$ TLRs. Adapted from Ref. \cite{LiShengNAABC2020}.}
\label{Fig ABCLattice}
\end{figure}

These two issues have been addressed in our recent work in Ref. \cite{LiShengNAABC2020}, where we derived explicitly that the PFC method can be extended to synthesize $\mathrm{U(2)}$ non-Abelian gauge fields in coupled TLR lattices, as shown in Fig. \ref{Fig ABCLattice}. The detailed scheme is similar to the described PFC scheme for the Abelian gauge field. For the lattice sites, we merely need to incorporate the $\lambda$ modes of the TLRs into consideration such that the $\lambda/2$ and the $\lambda$ modes of the individual TLRs play the role of the pseudo-spin components. As the lattice sites become multi-component, the linking variables $U_{{\mathbf{r^\prime}}, {\mathbf{r}}}=\exp \left[{i} \int_{\mathbf{r}}^{\mathbf{r^\prime}}\mathrm{d}\textbf{x} \cdot\textbf{A}(\textbf{x})\right]$ become matrix-valued as the non-Abelian background potential $\mathbf{A}\left( \mathbf{x} \right)$ is matrix-valued. The establishment of $\mathbf{A}\left( \mathbf{x} \right)$ in turn becomes the establishment of each of the hopping branch defined by the matrix elements of $U_{{\mathbf{r^\prime}}, {\mathbf{r}}}$. Thus, the generalization of the PFC method to the synthesization of the non-Abelian gauge field is straightforward: for TLRs coupled by common SQUIDs, we only need to guarantee that their eigenfrequencies are far off resonant, see Fig. \ref{Fig ABCLattice}(d), such that a multi-tone modulation can be applied to manipulate each hopping branch in the linking variables independently.

The physics of Aharonov-Bohm (AB) caging in a periodic 1D rhombic lattice is sketched in Fig. \ref{Fig ABCLattice}(a) \cite{Longhi:14,MukherjeeABCRealization2018PRL,Di2018Nonlinear}, with its TLR lattice realization shown in Fig. \ref{Fig ABCLattice}(b). Here each lattice site consists of $N$ (pseudo)spin modes. When exposed to an $\mathrm{U(N)}$ background gauge field $\mathbf{A}$, the Hamiltonian of the lattice takes the form
	\begin{equation}
	\label{Eqn RhombicHamiltonian}
	H=-J \sum_{\left\langle n\kappa, m\beta\right\rangle}\kappa_{n}^{\dagger} U_{{n}\kappa, {m}\beta} \beta_{m},
	\end{equation}
where $\kappa_{n}=[\kappa_{n,1},\kappa_{n,2},\cdots,\kappa_{n,N}]^T$ with $\kappa=A,B,C$ is the vector of the annihilation operators of the $\kappa$th site in the $n$th unit-cell, and $U_{{n}\kappa, {m}\beta}=\exp \left[{i} \int_{[m, \beta]}^{[n, \kappa]}\mathrm{d}\textbf{x} \cdot\textbf{A}(\textbf{x})\right]$. We then define the non-Abelian AB caging by the nilpotency of interference matrix
	\begin{equation}
	\label{Eqn NAABCCondition}
	I=\frac{1}{2}\left(U_{2}U_{1}+U_{4}U_{3}\right),
	\end{equation}
with $U_{1}$, $U_{2}$, $U_{3}$, and $U_{4}$ being the rightward link variables labeled in Fig. \ref{Fig ABCLattice}(a). For the Abelian situation $N=1$,  AB caging coincides with the $\pi$ magnetic flux penetration in each loop of the lattice. In this situation, a particle initially populated in the site $[n,A]$ cannot move to sites further than $[n \pm 1, A]$ due to the destructive interference of the two up and down paths shown in Fig. \ref{Fig ABCLattice}(a). For the non-Abelian situation $N > 1$, the matrix feature of $\mathbf{A}$ offers much more rich physics: A matrix-formed nonzero interference matrix $I$ can be nilpotent. To illustrate this idea, let us take an $\mathrm{U(2)}$ design as the minimal realization of the proposed non-Abelian AB caging. The link variables are set as
	\begin{equation}
	\label{Eqn U2NAABC}
	U_{1}=U_{4}=\left[\begin{matrix}
	1&0\\0&1
	\end{matrix}\right],
    U_{2}=\left[\begin{matrix}
	0&1\\1&0
	\end{matrix}\right],U_{3}=\left[\begin{matrix}
	0&1\\-1&0
	\end{matrix}\right],
	\end{equation}
such that a nilpotent
	\begin{equation}	
    \label{Eqn U2Interference}
	 I=\frac{1}{2}(U_2U_1+U_4U_3)=\left[\begin{matrix}
	 0&1\\0&0\end{matrix}\right],
	\end{equation}
can be achieved. That is, an excitation in the $[n,A,1]$ mode can move only to the $[n+1,A,2]$ mode in a  rightward way, but then it can move to neither the $[n+2,A]$ nor the $[n-1,A]$ sites. Similarly, an excitation  in the $[n,A,2]$ mode can move to the $[n-1,A,1]$ mode, but it cannot arrive the $[n-2,A]$ and $[n+1,A]$ sites. This is due to the fact that the non-Abelian feature of the synthetic  gauge field leads to exotic asymmetric spatial configuration of excitations, which is determined by both the nilpotent power   $I$ and the initialization of the photon. These exotic new features stem from the non-Abelian nature of $\mathbf{A}$ and have no Abelian analog.

\subsection{TBT in 2D}

\begin{figure}[tb]
\begin{center}
    \includegraphics[width=0.45\textwidth]{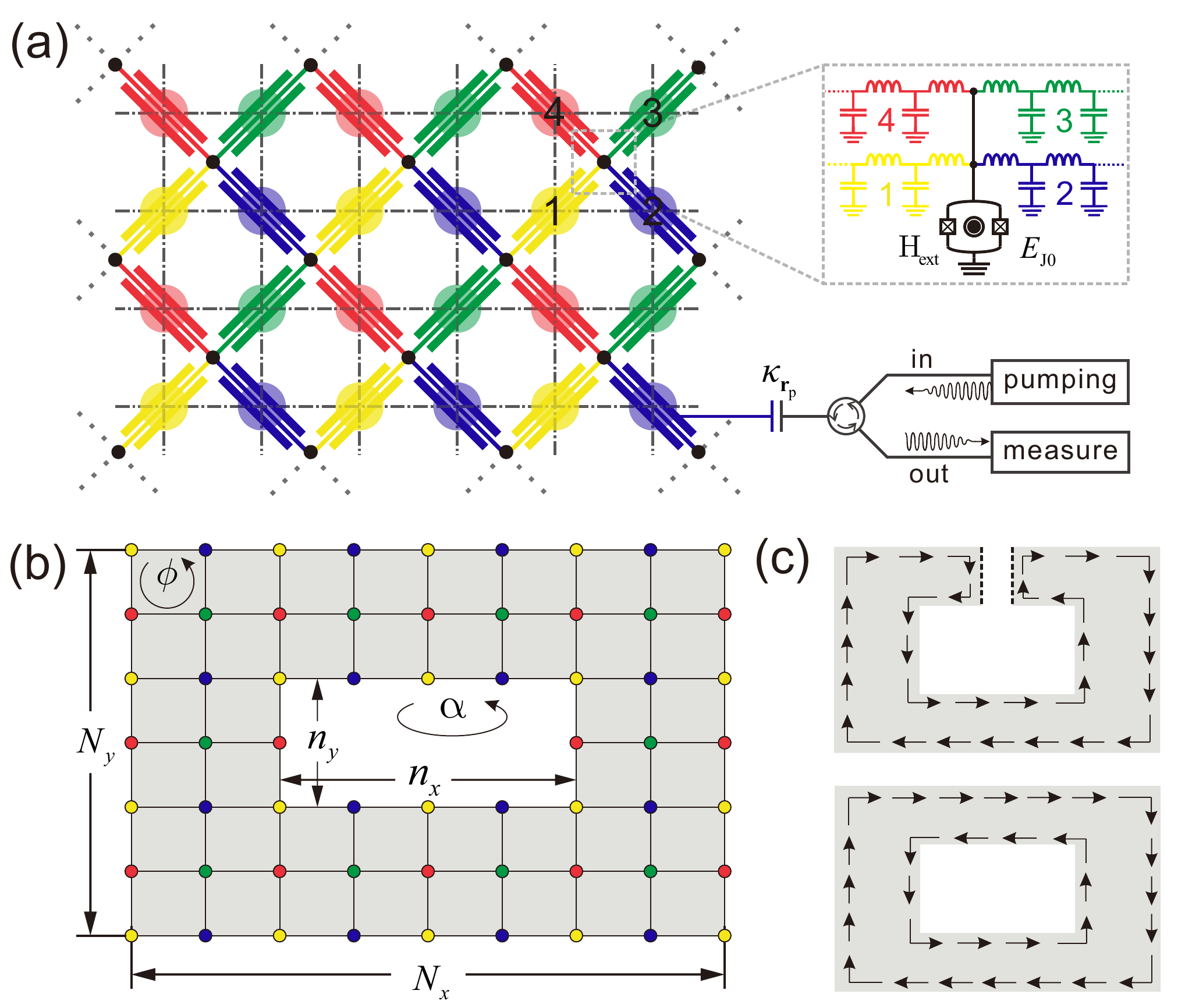}
\end{center}
\caption{ (a) Sketch of the TLR square lattice simulator of IQHE. The four TLRs building the lattice are labelled by the four different colors and the grounding SQUIDs are labelled by the black dots. The effective hopping between the TLRs is established through the PFC modulation of the SQUIDs. (b) Geometric configuration of the proposed lattice.  (c) Chiral property of the ESMs located at the inner and outer edges. The lattice sketched in (b) can be regarded as the gluing of a simply-connected plane by its two sides (the upper panel). Such gluing process can intuitively interpret a variety of opposite properties of the inner and the outer ESMs, with the opposite chiral property being the prime example (the arrows in the lower panel). Adapted from Ref. \cite{WangYPNPJQI2016}. }
\label{Fig SquareLattice}
\end{figure}

With the developed technique of synthesizing gauge fields by the PFC method, we further consider the quantum simulation of integer QHE and the consequent measurement of integer topological quantum number in a  2D photonic lattice \cite{WangYPNPJQI2016}.  For this aim, we propose a square lattice consisting of TLRs with four different frequencies, placed in a square lattice form and grounded by SQUIDs at their common ends, as shown in Fig.~\ref{Fig SquareLattice}(a). Similar to Ref. \cite{WangYPChiral2015}, the eigenfrequencies of the four kinds of TLRs are specified as $\left[\omega_{\mathrm {0}},\omega_{\mathrm{0}}+\Delta,\omega_{\mathrm{0}}+3\Delta,\omega_{\mathrm{0}}+4\Delta\right]$, respectively. In each of the grounding SQUIDs, a developed three-tone PFC pulses is applied, such that the hopping strength $J$ and hopping phase of every branch can be independently controlled. Such \textit{in situ} tunability can  in turn lead to the synthesization of the following  artificial gauge field for microwave photons on the lattice
\begin{align}
\mathbf{A}&=\left[ 0, A_y(x), 0 \right], \notag\\
\mathbf{B}&=B\mathbf{e}_z=\left[ 0, 0, \frac{\partial}{\partial x} A_y(x) \right].
\end{align}
For the aim of measuring the integer topological invariants of this system, a nontrivial ring geometry is further endowed to the TLR lattice, that is, we consider an $N_{x} \times N_{y}$ square lattice with an $n_{x} \times n_{y}$ vacancy at its middle, as shown in Fig.~\ref{Fig SquareLattice}(b). An uniform magnetic flux $\phi$ in each plaquette of the lattice and an extra magnetic flux $\alpha$ at the central vacancy can be penetrated through the careful setting of the hopping phases of the hopping branches. This configuration closely mimics the previously discussed Laughlin cylinder shown in Fig. \ref{Fig Laughlinring}.

Following the bulk-edge correspondence principle \cite{LaughlinBEC1981PRB}, the ESMs of the lattice provides an efficient handle to probe the topological invariants of the bulk band. Explicitly speaking, the ESMs located in the band gap between the $h$th and the $(h+1)$th bands can be characterized by the topological winding number $w_{h}$, which can be related to the Chern number $\mathcal{C}_{h}$ of the $h$th band as
\begin{equation}
\label{Eqn ChernNumber}
\mathcal{C}_{h}=w_{h}-w_{h-1}.
\end{equation}
Therefore, the measurement of the Chern numbers $\mathcal{C}_{h}$ of the bulk bands is equivalent to the measurement of the winding numbers $w_{h}$ of the ESMs. For the rational magnetic flux situation $\phi/2\pi=p/q$ with $p,q$ being co-prime integers, the topological winding numbers $w_h$ of the ESMs are given by the Diophantine equation
\begin{equation}
\label{Eqn diophantine}
h=s_{h}q+w_{h}p, |w_{h}| \leq q/2 ,
\end{equation}
with $h$ being the gap index and $w_{h}, s_{h}$ being integers.

\begin{figure}[tb]
\begin{center}
\includegraphics[width=0.48\textwidth]{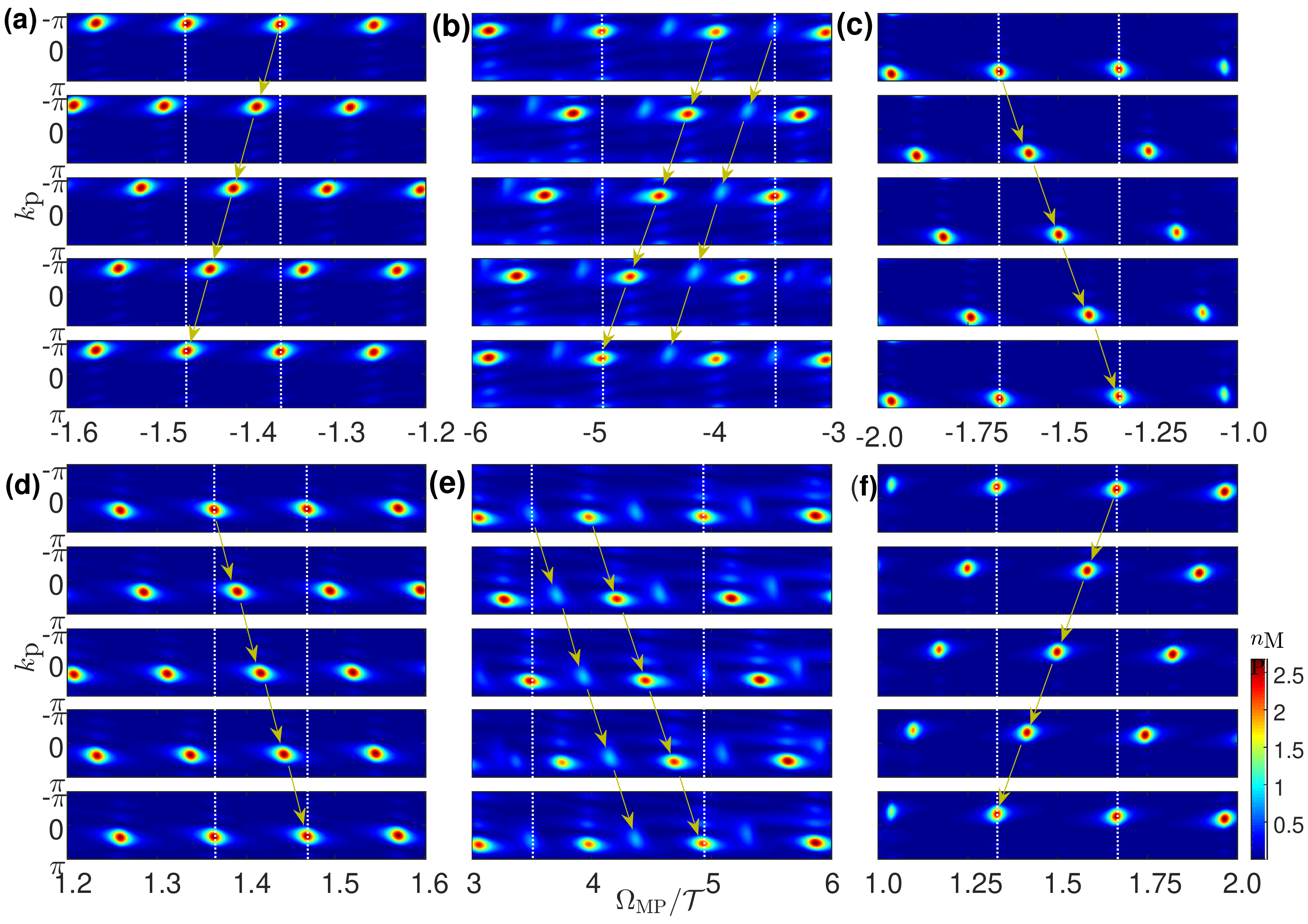}
\end{center}
\caption{\label{Fig WindingnumberLattice} Laughlin pumping in the proposed square lattice. Here the uniform magnetic field is set as $\phi/2\pi=1/4$ for  (a),  (c), (d), and  (f), and $\phi/2\pi=1/5$ for  (b) and  (e). In each of the subfigures, the panels correspond to $\alpha/2\pi=0$, $1/4$, $1/2$, $3/4$ and $1$ from top to button, respectively. For  (a), (b), (d), (e), the pumped ESMs are located at the outside edge of the lattice, while for (c) and (f)  the pumped ESMs are located at the inner side edge. Adapted from Ref. \cite{WangYPNPJQI2016}.}
\end{figure}

Since the spatial configuration of the proposed lattice is equivalent to the Laughlin cylinder shown in Fig. \ref{Fig Laughlinring}, we can realize the adiabatic Laughlin pumping of the ESMs through controlling the central vacancy flux $\alpha$. By increasing $\alpha$ monotonically from $0$ to $2\pi$, we can observe that the spectrum of the lattice varies and  returns finally to its original form, with an integer number of ESMs been transferred. This integer is exactly the winding number $t_h$ of the ESMs. In Figs.~\ref{Fig WindingnumberLattice}(a) and \ref{Fig WindingnumberLattice}(d),  the ESMs spectrum versus $\alpha$ is numerically simulated for $p/q=1/4$. The guidance of the dashed white lines and the solid yellow arrows clearly indicates the movements of the ESM peaks in the $1$st and $3$rd gaps, which is in agreement with the calculated topological winding numbers $w_{1}=-w_{3}=1$. A similar situation $\phi/2\pi=1/5$ is also calculated and shown in Figs.~\ref{Fig WindingnumberLattice}(b) and \ref{Fig WindingnumberLattice}(e), where we observe that  the ESMs in the $2$nd and $3$rd gaps cross two peaks with opposite moving directions during the whole pumping process. This result is in accordance with the calculated $w_{2}=-w_{3}=2$. In addition, Figs.~\ref{Fig WindingnumberLattice}(c) and  \ref{Fig WindingnumberLattice}(f) indicate that the ESMs in the inner and outer edges take the opposite moving directions during the pumping process. In summary, this measurement scheme of the topological invariants can be summarized as follow: during the Laughlin pumping process,  we only need to follow the movement of the ESM peaks, as ESMs in the $h$th gap will move by $|w_{h}|$ peaks, with the moving direction determined by the sign of $w_{h}$. From the obtained $w_{h}$, we can directly calculate the Chern numbers $\mathcal{C}_{h}$ following the relation Eq. (\ref{Eqn ChernNumber}).

\begin{figure}[tb]
\begin{center}
\includegraphics[width=0.47\textwidth]{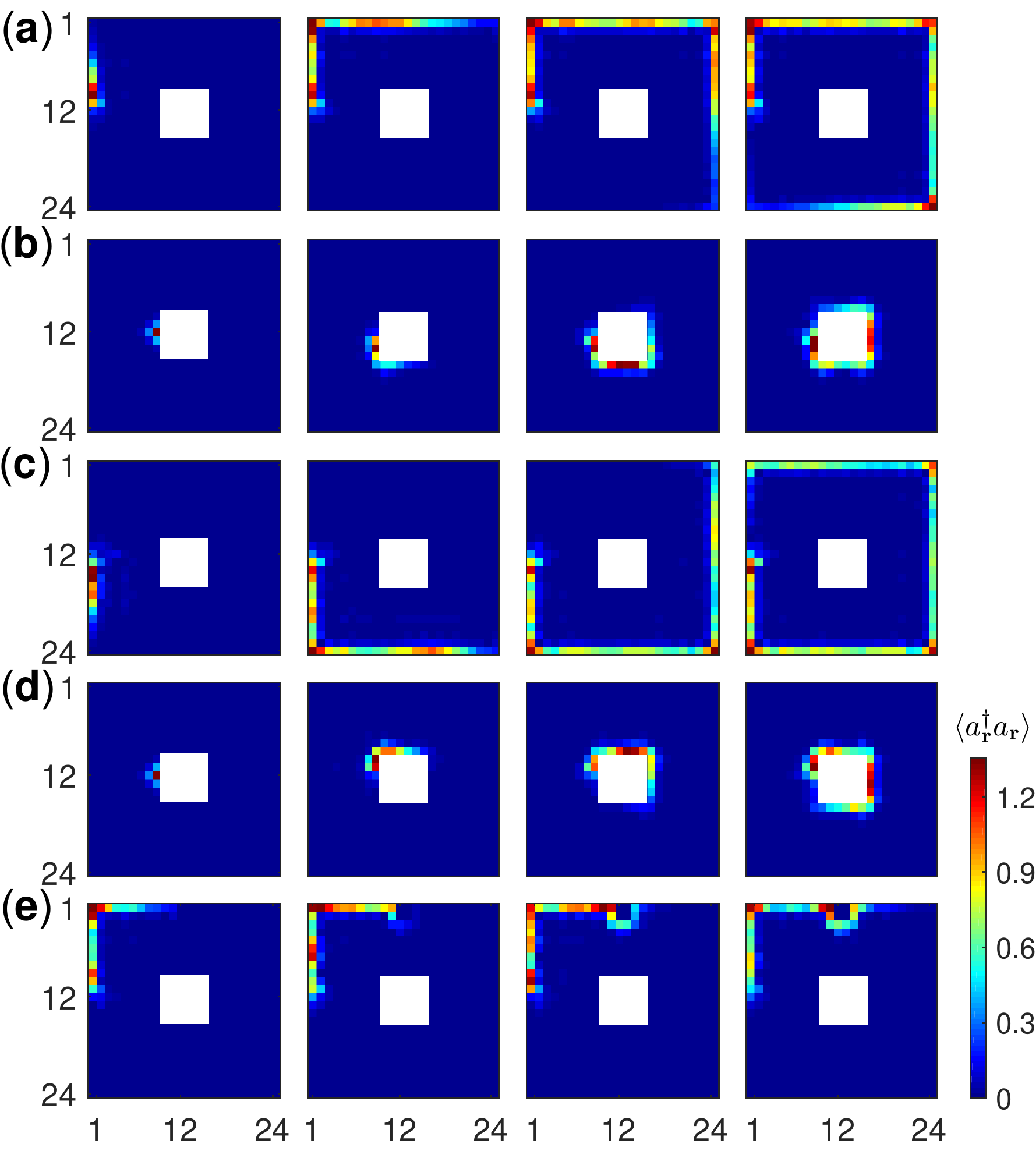}
\end{center}
\caption{Chiral photon flow dynamics on the TLR lattice. The pumping sites are selected as $\mathbf{r}=(1,13)$ for  (a), (c), and  (e), and $\mathbf{r}=(9,13)$ for (b) and  (d). In addition, we set the imposed synthetic magnetic field as $\phi/2\pi=1/4$ and $\alpha/2\pi=0$; $\Omega_{\mathrm{SP}}/J$ is chosen as $[-1.76$, $-1.97$, $1.47$, $1.97$, $-1.75]$ for  (a)--(e), respectively. The times of the panels from left to right are arithmetic progressions with the first terms $J_1/2\pi=[6$, $1$, $6$, $1$, $15]J^{-1}$ and the common differences $\Delta J/2\pi=[13$, $5$, $13$, $5$, $2.5]J^{-1}$ for (a)--(e), respectively. In the calculation of  (e) we have incorporated the effect of lattice disorder and defect. Adapted from Ref. \cite{WangYPNPJQI2016}.}
\label{Fig LatticeChiral}
\end{figure}

The extension of the discussed chiral photon flow on the necklace is the coherent chiral photon flow dynamics of the ESM on the lattice. Such investigation can offer an intuitive insight into the chiral property of the ESMs. We assume that we initialize the lattice in its ground state and then add a driving pulse on an edge site $\mathbf{r}_\mathrm{p}$ with the pumping frequency $\Omega_{\mathrm{SP}}$ being an ESM eigenfrequency. Then we numerically simulate the time evolution of the lattice and sketch several screenshots of the photon flow dynamics in Fig.~\ref{Fig LatticeChiral}, from which we can observe the unidirectional photon flows around the edge and the directions of the chiral photon flow are determined by both $\Omega_{\mathrm{SP}}$ and $\mathbf{r}_\mathrm{p}$. If we choose $\mathbf{r}_\mathrm{p}$ on the outside edge, we can find out that the chiral photon flow is clockwise /counterclockwise if we choose $\Omega_{\mathrm{SP}}$ in the 1st/3rd gap (Figs.~\ref{Fig LatticeChiral}(a) and \ref{Fig LatticeChiral}(c)). The different chirals of the ESMs in different energy gaps can be understood by regarding the chiral photon flow as a macroscopic rotating spin \cite{HafeziNatPhoton2013}: if we place such a rotating spin in a magnetic field $\mathbf{B}$, its energy will split according to its spinning direction.  Moreover, from Figs.~\ref{Fig LatticeChiral}(a) and \ref{Fig LatticeChiral}(b) (also Figs.~\ref{Fig LatticeChiral}(c) and \ref{Fig LatticeChiral}(d)) we can observe that the inner and outer ESMs in the same gap have opposite chiral properties. This can be explained by the spatial topology of the proposed lattice. As shown in the upper panel of Fig.~\ref{Fig SquareLattice}(c),  if we tear the lattice apart, we can get a simply-connected plane with only one connected edge. Meanwhile, the inverse of the tearing, that is, the gluing of the two sides marked by dashed lines,  cancels the ESM flow (marked by the arrows) on the glued sides,  and results in finally  two closed circulations located at the inner and outer edges with opposite circulating directions. As each of the hopping branches can be tuned in a site resolved way by the proposed PFC formalism, we can expect the future experimental demonstration of this ``tearing-and-gluing'' process.

The influence of disorder and defect on the chiral flow is also calculated. In this numerical simulation, we add Gaussian distributed diagonal and off-diagonal disorder with $\sigma(\delta\omega_\mathbf{r})=\sigma(J_\mathbf{r^{\prime}r})=0.05 J$. In addition, we place a $2 \times 2$ hindrance on the upper outer edge with $\delta\omega_{(12-13,23-24)}/J=30$. In this situation, the topological robustness of the ESMs is clearly verified by the survival of the chiral photon flow displayed in Fig.~\ref{Fig LatticeChiral}(e).

\subsection{Extension: Parametric amplification}

A direct extension of the proposed PFC method is the parametric amplification (PA) process. Compared with the described synthetic gauge field issue, this process is unique for photonic systems. We come back to Eq. (\ref{equ:drivingsimple}). Now we modulate $g_{12}(t)$ by the summation tone of the two resonators:
\begin{equation}
g_{12}(t)=2J\cos[(\omega_{1}+\omega_{2})t]. \label{equ:PAdriving}
\end{equation}
Then we can get an effective nondegenerate parametric amplification  Hamiltonian
\begin{align}
{H}^{12}_{\mathrm{eff}}&=e^{iH_{0}t}{H}^{12}_{\mathrm{ac}}(t)e^{-iH_{0}t}\notag\\
&\approx g_{12} a_{1}^{\dag
}a_{2}^{\dagger}+\mathrm{h.c.}, \label{equ:PAHamiltonianchiral0}
\end{align}
which takes the similar form of the fermionic p-wave pairing \cite{Kitaev}. From this point of view, the road map of simulating  one dimensional fermionic topological superconductor models is clear: The lattice sites are built by  hard-core boson,  e.g. TLRs with strong nonlinearity or superconducting transmon qubits such that the fermionic statistics can be induced by the Jordan-Wigner transformation \cite{FermionizedPhoton,ImamogluMajoranaPRL2012}, while the hopping and pairing terms between the lattice sites are induced by the PFC and parametric  amplification modulation of the coupling grounding SQUIDs, respectively. With further modified dispersive amplification modulation method, this architecture can be even exploited to simulate spinful fermionic models. Actually, this idea has already been illustrated  in Ref. \cite{HuWangDIII2017}, where we proposed a 1D  hard-core boson lattice built by superconducting transmon qubit as a  faithful simulator of the following fermionic time-reversal invariant DIII model \cite{ZhaoWang-MF}
\begin{eqnarray}
H_{\mathrm{DIII}}& =& -\mu ( c_{j,\epsilon }^{\dagger }c_{j,\epsilon }-1) \notag\\
&&-\sum_{j\epsilon }(g c_{j,\epsilon }^{\dagger }c_{j+1,\epsilon}
+i\Delta c_{j+1,\epsilon }c_{j,\bar{\epsilon }}+\mathrm{h.c.}), \label{Eq DIIIModel}
\end{eqnarray}%
where $c_{j,\epsilon }^{\dagger }$ is the creation operator of the spin-$\epsilon$ fermion on the $j$th site, $g$, $\mu$, $\Delta $ denote the real-value hopping strength, the chemical potential, and the amplitude of p-wave pairing parameter. The nontrivial coupling between the lattice sites is induced through the inductive connection between transmon qubits \cite{KochTransmonPRA2007, SchreierPRB2008, MartinisCoupling1, MartinisCoupling2}. In particular, we developed a generalized  parametric amplification approach to facilitate the exotic $U(1)$ gauge factor emerged in the  Jordan-Wigner bosonization. Furthermore, we proposed that the odd-parity Kramers doublet ground states can be used as the basis of topological qubits,  and with these topological bases the universal topological quantum gates can be implemented in principle.

\section{TBT with superconducting qubits}

\subsection{TBT in 1D}
With the above tunable coupling among qubits, we are readily in the position to simulate topological quantum phases on superconducting quantum circuits. The first example is the  1D SSH model with the Hamiltonian  of  Eq. (\ref{Eqn SSHHamiltonian}), introduced in Section \ref{SSH} and it can be realized by parametrically tuning the inter-qubit coupling strength in Eq. (\ref{eq:H}) in the required dimerized  form. Note that, although the Hamiltonian in Eq. (\ref{eq:H}) is of the bosonic nature, in the single excitation subspace, its topology is equivalent to that of the SSH model, which has two different topological phases with  different winding numbers.  Specifically, when the qubit couplings are tuned into $g_A < g_B$ ($g_A > g_B$) configuration, with $g_A \equiv g'_{2j+1}$ and $g_B \equiv g'_{2j+2}$  in Eq. (\ref{eq:H}), the winding number will be 0 (1), and then the system is in a topologically trivial (nontrivial) state. Then, by monitoring the quantum dynamics of a single-qubit excitation in the chain,  the associated topological quantities  can be inferred \cite{CD}.

\subsection{TBT in 2D}
The gauge effect is essential in investigating novel phenomena in modern physics, e.g., for exploring exotic quantum many-body physics. In this subsection, we review two schemes to synthesize gauge field by introducing ac modulation on superconducting circuits, and the
motivation is to obtain an extremely strong effective magnetic field for   bosons, which is hard  in conventional solid-state systems. While  these pioneering examples are associated with the chiral effect,  novel quantum many-body properties induced by the interplay of the synthetic gauge field, bosonic hopping, and interaction of bosons are also deserve further exploration.

\begin{figure}[tb]
\begin{center}
\includegraphics[width=8cm]{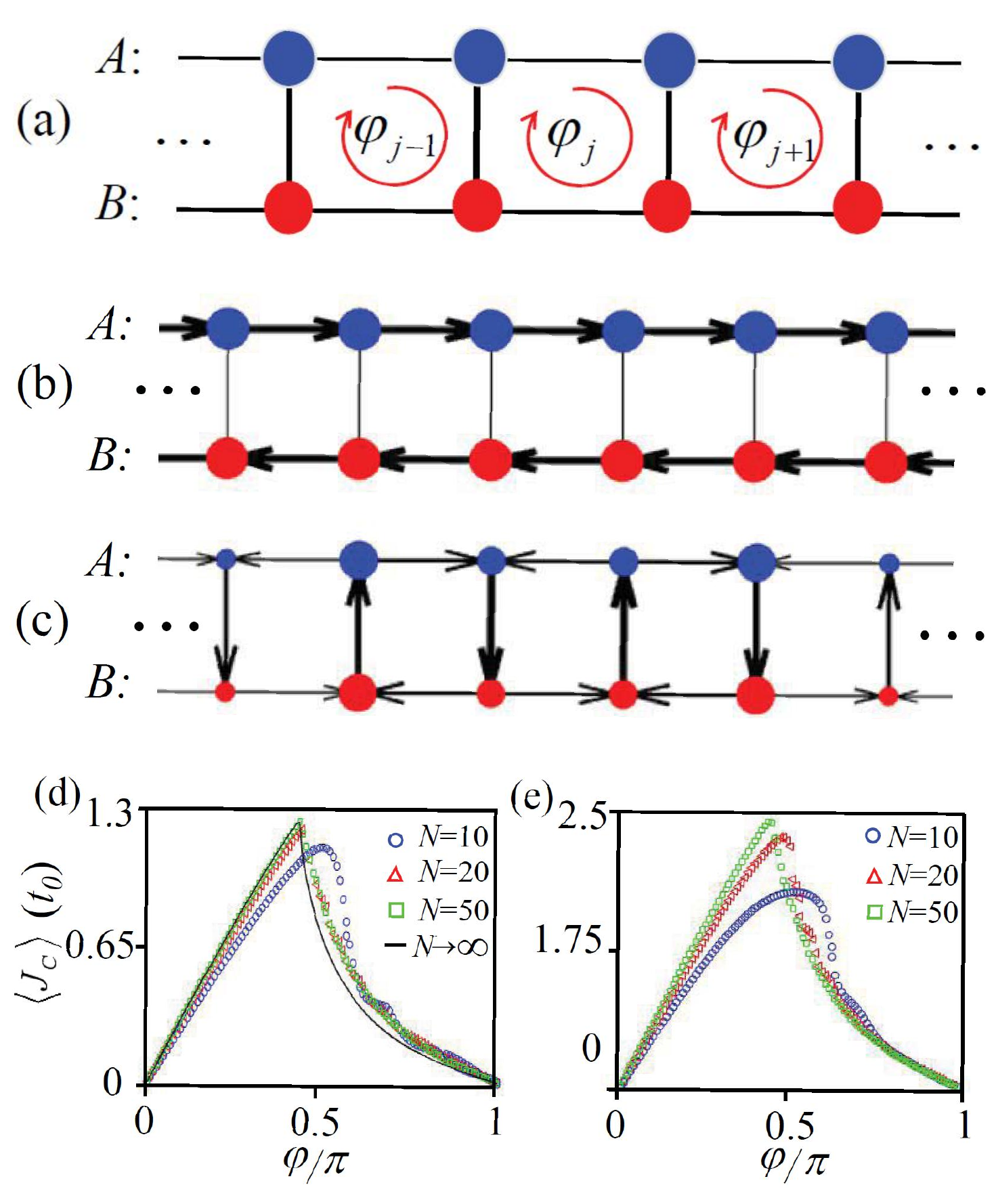}
\end{center}
\caption{The chiral currents for a two-leg quantum circuit. (a) Schematic diagram of the two-leg lattice model with an synthetic magnetic flux per plaquette. The blue and red solid circles represent transmon qubits at the $A$ and $B$ legs. The chiral currents between  neighboring sites in the single-excitation case for (b) $\protect\varphi =0.1\protect\pi $ and (c) $\protect\varphi =0.9\protect\pi $ with $N=50$, where the thicknesses of the arrows represent their strengths and the sizes of the circles denote their local densities. The ground-state chiral current  as a function of $\protect\varphi $ for (d) single- and (e) two-excitation cases. The solid line in (d) is plotted from the analytical result. Reproduced from Ref.   \cite{twoleg}. }
\label{current}
\end{figure}

We consider a two-leg superconducting circuits consisting of superconducting transmon qubits, as shown in Fig.~\ref{current}(a). Extending the parametric coupling in Eq. (\ref{eq:H}) to this two-leg case, the considered model  Hamiltonian here is \cite{twoleg}
\begin{eqnarray}
\hat{H} &=&\underset{\nu j}{\sum }\left( t_{\nu j}e^{i\tilde{\varphi}_{\nu
j}}\hat{a}_{\nu \left( j-1\right) }^{\dag }\hat{a}_{\nu j}+\text{H.c.}\right)
\notag \\
&&+\underset{j}{\sum }\left( \tilde{t}_{j}e^{i\tilde{\varphi}_{Bj}}\hat{a}%
_{Aj}^{\dag }\hat{a}_{Bj}+\text{H.c.}\right) ,  \label{H2}
\end{eqnarray}
where $j$ denotes the number of the rung, $\nu \in \{A, B\}$ labels the leg, $\hat{a}_{\nu j}^{\dag }$ ($\hat{a}_{\nu j}$) is the creation (annihilation) operator for the $j$th site on the $\nu $th leg, $\tilde{\varphi}_{\nu j}=\left( -1\right) ^{j+1}\varphi _{\nu j}+\pi /2$, $t_{\nu j}=g_{\nu j}J_{0}(\eta _{\nu j-1})J_{1}(\eta _{\nu j})$, and $\tilde{t}_{j}=\tilde{g}_{j}J_{0}(\eta _{Aj})J_{1}(\eta _{Bj})$ with   $g_{\nu j}$ is the original static hopping strength between the nearest-neighbor sites along the leg $\nu $, $\tilde{g}_{j}$ is the static interleg coupling strength at the rung $j$.
In this two-leg lattice model, the  non-zero magnetic flux case supports  a chiral current follow within the lattice, which is  plotted \cite{twoleg} in Fig.~\ref{current}. For simplicity, we have set $\eta _{\nu j}=\eta $ and $g_{\nu j}=\tilde{g}_{j}=g$, and thus $t_{\nu j}=\tilde{t}_{j}=t_{0}=gJ_{0}(\eta )J_{1}(\eta )$. Meanwhile, setting $\tilde{\varphi}_{Aj}=-\tilde{\varphi}_{Bj}=\varphi /2$,  the synthetic magnetic flux will be $\varphi _{j}=\varphi$.  The chiral currents between any  neighboring sites are plotted for (b) $\varphi =0.1\pi $\ and (c) $\varphi =0.9\pi$ cases with $N=50$. While in Fig.~\ref{current}(d),  the ground-state chiral current   is plotted as a function of $\varphi $, where as the increasing of $\varphi $, the current  increases to a maximal value at the critical
point $\varphi _{c}$ and then decreases. This indicates a quantum phase transition appears, i.e., from the Meissner phase to the vortex phase \cite{MP15}, which survivals in a finite size system, and thus is readily  observable in current experimental setups. In Fig.~\ref{current}(e),  the ground-state chiral current for the two-excitation case is plotted, where shows similar properties as that of the single-excitation case.


\section{TBT with Dressed state systems}
The above discussed  quantum simulation examples on superconducting circuits are limited to the spinless cases, as it is difficult to engineer the spin degree of freedom in either TLR or qubit systems. To simulate spinful systems, one needs to introduce more degrees of freedom. One of the possible platform is the TLR-qubit coupled system, a typical circuit QED setup \cite{sqc4}, and the dressed-state or polariton of which can mimic the spin degree of freedom \cite{xue}. In this platform, the synthetic polaritonic SOC and Zeeman field can be induced simultaneously with fully anin situ tunability \cite{xueso}, and thus it provides a flexible platform to explore topological physics.

\begin{figure}[tbp]\centering
\includegraphics[width=8cm]{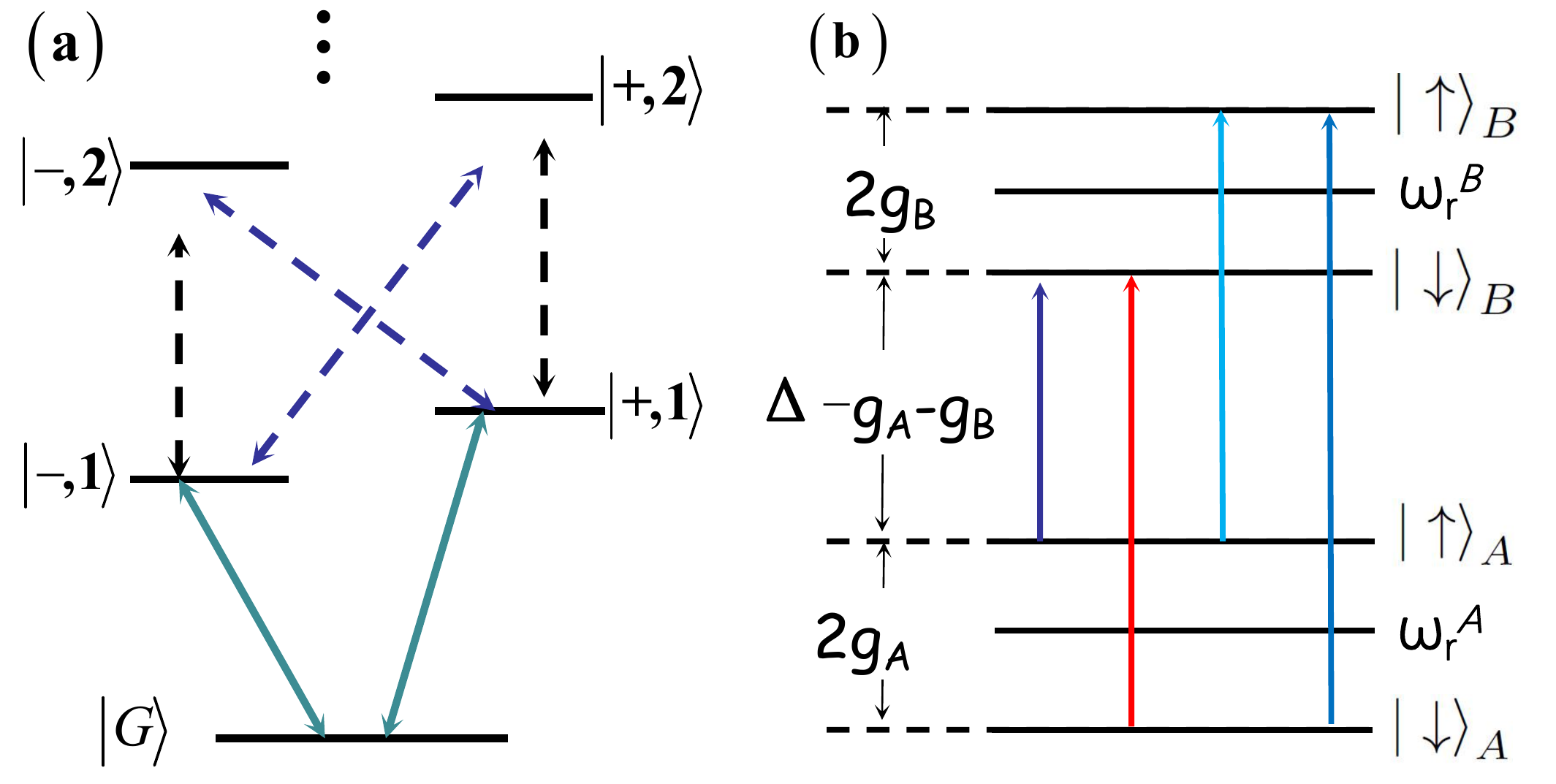}
\caption{Polariton states and the two-polariton transitions. (a) The eigen energy spectrum of the JC coupled circuit QED system. Under approximate parameters of a two-tone driven field, the lowest three levels, of the V-configuration, can be addressed to realize single-polariton-qubit manipulations.  (b) The  transitions of two-polariton systems with $\Delta> g_A+g_B$.}
\label{figpolariton}
\end{figure}

\subsection{The Dressed-state qubits}

We begin with implementing the dressed-state on a conventional circuit QED setup, where a superconducting qubit of the transmon type is capacitively coupled to a 1D TLR. The interaction Hamiltonian is in the Jaynes-Cummings (JC) form of \cite{sqc4}
\begin{eqnarray}\label{jc}
H_{\text{JC}} = \omega_{\text{a}} |1\rangle\langle1| +\omega_{\text{r}} a^{\dagger}a +g(a\sigma^{+} + a^{\dagger} \sigma^{-}),
\end{eqnarray}
where $\omega_{\text{a}}$ and $\omega_{\text{r}}$ are the frequencies of qubit and cavity, respectively; $|n\rangle_{\text{r}}$ labels the Fock state of the TLR, the ground state of Hamiltonian in Eq. (\ref{jc}) is $|G\rangle=|g\rangle|0\rangle_{\text{r}}\equiv|g,0\rangle$ with its eigenvalue is set to be $E_{G}=0$. As shown in Fig. \ref{figpolariton}(a), for $n\geq1$, due to the  TLR-qubit interaction,  the two-fold eigenstates for a certain \emph{n} are
\begin{subequations}
\begin{eqnarray}
{|-,n\rangle}&=\cos\alpha_{n}|g,n\rangle-\sin\alpha_{n}|e,n-1\rangle,\\
{|+,n\rangle}&=\sin\alpha_{n}|g,n\rangle+\cos\alpha_{n}|e,n-1\rangle,
\end{eqnarray}
\end{subequations}
and corresponding eigenvalues are
\begin{eqnarray}\label{E} 
E_{n,\pm} = n\omega_{\text{r}}+{1 \over 2} \left(\delta\pm\sqrt{\delta^2+4ng^2}\right),
\end{eqnarray}
where $\tan(2\alpha_{n})=2g\sqrt{n}/\delta$ with the detuning $\delta=\omega_{\text{r}}- \omega_{\text{a}}$. Note that the transition frequencies among eigenstates are different, and thus selective transition between two target states may be achieved.  The three lowest eigenstates are $|G\rangle$, $|-,1\rangle$, and $|+,1\rangle$, which form a V-type artificial atom,  as shown in Fig. \ref{figpolariton}(a).  For the sake of simplicity, we denote the polaritonic states $|+,1\rangle$ and $|-,1\rangle$ as  $|\uparrow\rangle$ and $|\downarrow\rangle$ in the following,   mimic a spin 1/2 system, or a spin qubit. Compared with conventional superconducting qubits, this combined definition of qubit has better stability against low-frequency noises \cite{xueepjd}.

\begin{figure}[tb]
\centering
\includegraphics[width=7cm]{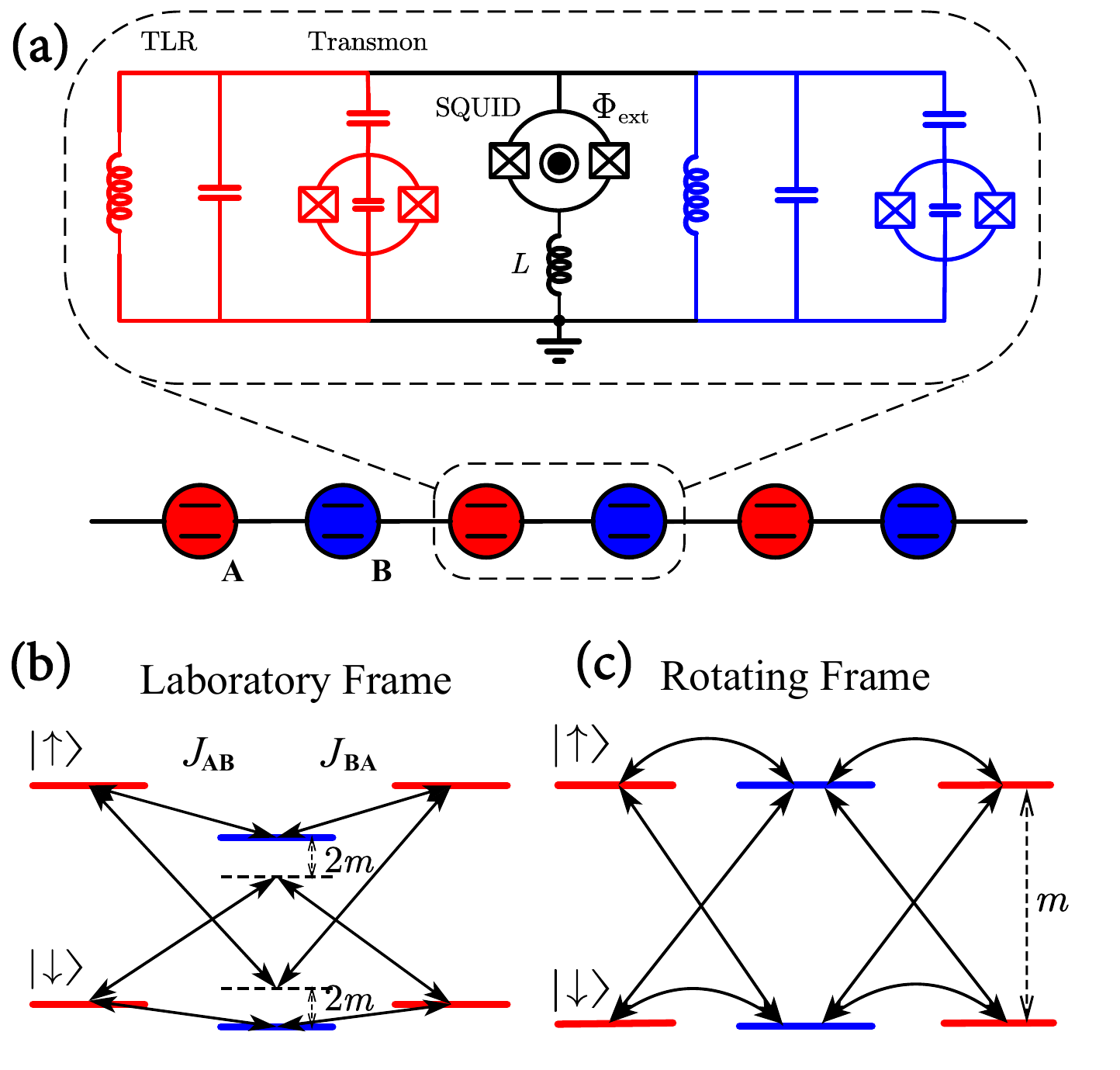}
\caption{Implementation of 1D spin-1/2 lattice models on a superconducting quantum circuit.  (a) The ``spin-1/2" lattice with A- (red) and B-type (blue) of unit cells, arranged alternately.  The zoom-in figure details the equivalent superconducting circuits of the unit.   (b) The resonant spin-preserved and detuned spin-flipped  couplings of different spin states. Since the alternative arrangement, two sets of driving, $J_{\text{AB}}$ and $J_{\text{BA}}$, are needed to preserve the translation symmetry. (c) In the rotating frame, and the levels and designed hoppings of both  A- and B-types of the unit cells can be treated as identical.  Adapted from Ref. \cite{xueso}.}		\label{figpolariton1d}
\end{figure}

\subsection{SOC for polariton spin}

The described parametric coupling scheme for the microwave TLRs can also be used here with the inclusion of a transmon qubit in each TLR. As the dressed   qubits are of the half-TLR plus half-transmon nature, the effective hopping among them can be  induced by only coupling the photonic component.  Here, we  present an example with two dressed   qubits case, i.e., A- and B-type TLR-qubit coupled system, which are different as they have different eigen-frequencies ($\omega^A_r$ and $\omega_r^B$) and coupling strengths ($g_A$ and $g_B$) in the JC Hamiltonian of Eq. (\ref{jc}). Setting $\delta=0$.  The energy spectrum and their transitions are shown in Fig. \ref{figpolariton}(b) when $\omega^B_r -\omega_r^A=\Delta> g_A+g_B$. The transitions can be induced with the parametric photonic coupling, and the addressing of each transition is obtained when the frequency of the coupling strengths between  TLRs, see Eq. (\ref{equ:drivingsimple}),  meet the resonant condition of a target transition \cite{xue}. Meanwhile, the energy difference of the four different hoppings are $2g_A$ or $2g_B$, which can be set to be much larger than the effective hopping strength between the two polariton states, to ensure the selective frequency addressing of each transition. Therefore, the fully tunable spin-preserved hopping  and SOC terms  can be induced in a same setup, and thus this element can naturally be scaled up into different types of lattices to study   virous SOC topological states.

\subsection{TBT in 1D}
We firstly proceed to present a spin-1/2 chain with adjustable SOC and Zeeman field that can be simulated with the above JC lattice \cite{xueso}. We  set the  A- and B-types units to be arranged in an alternate way, as shown in Fig. \ref{figpolariton1d}(a).  Then,   the energy differences of the four hopping  are also set to be much larger than the effective hopping strength to ensure the selective frequency addressing of the target transitions. Moreover, a detuning $2m$ to the spin-flipped transition tunes, as shown in Fig. \ref{figpolariton1d}(b),   can induce a tunable spin splitting $m$ for each cell, in the rotating frame, as shown in Fig. \ref{figpolariton1d}(c). In this way, neglecting the fast oscillating  terms, we obtain
\begin{equation} \label{eq.simu}
H'_{\text{JC}}=\sum_{l}^{N} m \bm{S^z}_{l}
+\sum_{l=1}^{N-1} \sum_{\epsilon, \bar{\epsilon}} \left(t_{0,l\epsilon\bar{\epsilon}} \text{e}^{\text{i}\varphi_{l\epsilon\bar{\epsilon}}}
\hat{c}^\dagger_{l,\epsilon} \hat{c}_{l+1,\bar{\epsilon}} +\text{h.c.}\right),
\end{equation}
where $\bm{S^z}_{l}=| \uparrow \rangle_l\langle \uparrow |- | \downarrow \rangle_l\langle  \downarrow  |$, $\hat{c}^\dagger_{l,\epsilon}=|\epsilon\rangle_l \, \langle G | $ is the creation operator for  spin-$\epsilon$ polariton state at the $l$th unit cell, and $t_{0,l\epsilon\bar{\epsilon}}$ is the effective hopping  strength. Therefore, the  Hamiltonian in Eq. (\ref{eq.simu}) simulates a 1D tight-binding lattice model consists of spin-1/2 particles, with $m$, $t_{0,l\epsilon\bar{\epsilon}}$ and $\varphi_{l\epsilon\bar{\epsilon}}$  being the corresponding effective Zeeman energy, coupling strength  and phase, which can respectively  be tuned  via  the deliberately chosen of the frequencies, amplitudes and phases of the external ac driving field. Note that although we used two types of unit cells, in an approximate  rotating frame and  by adjusting the two types of coupling parameters $J_{\text{AB}}$ and $J_{\text{BA}}$,  the translation symmetry in the model can still be preserved, and thus different sites can then be treat as identical.

\begin{figure}[tb]		
\centering
\includegraphics[width=8cm]{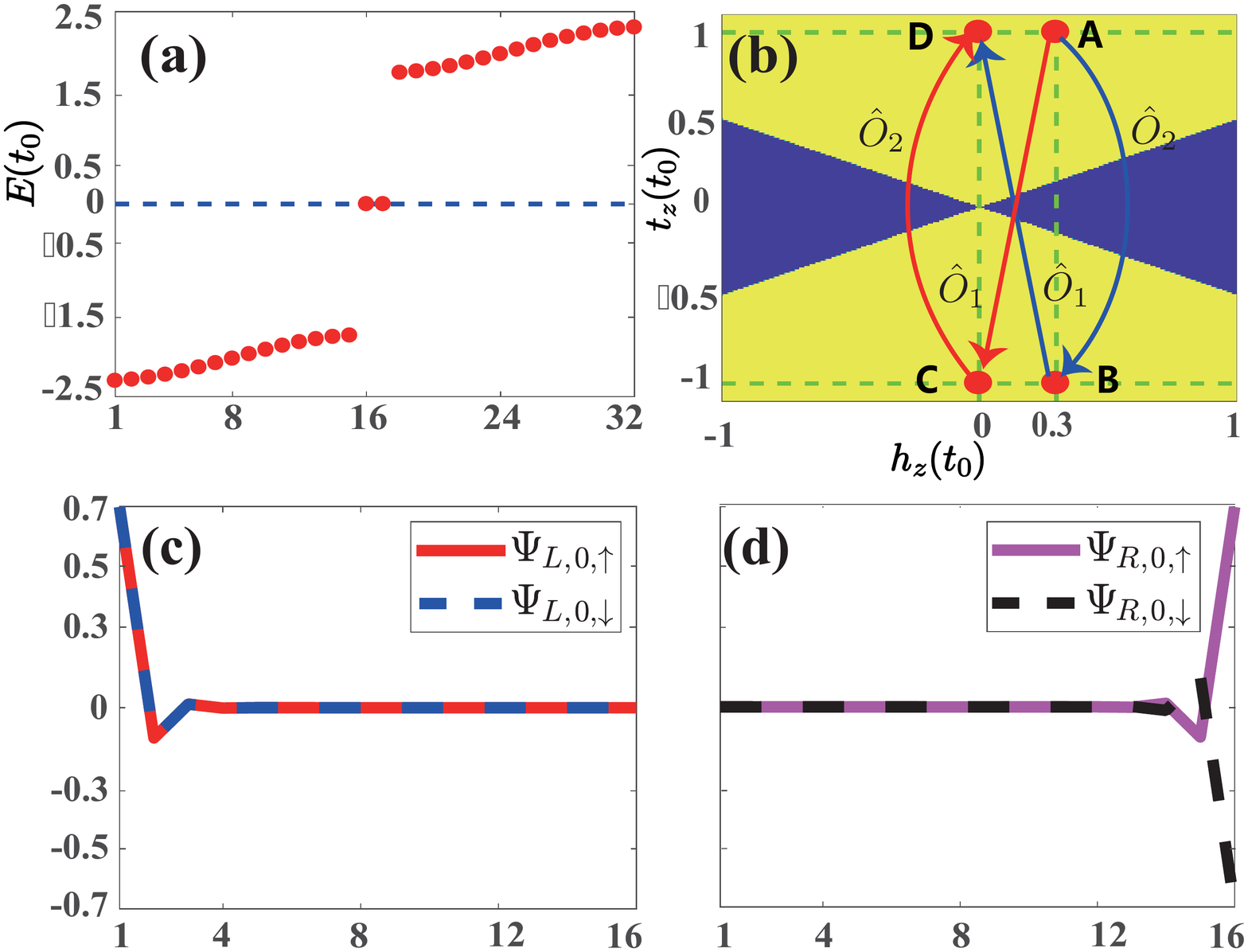}
\caption{Numerical results for a 1D lattice  with 16 unit-cells. (a) Spectrum of eigenenergies  in terms of an energy unit $t_0$ with $t_z/t_0=1$, $\Delta_0/t_0 = 0.99$, and $h_z/t_0 = 0.3$, where two ESMs appears.  (b) The Phase diagram of the chain with arbitrary $\Delta_0$. The yellow and dark blue regions denote the topologically nontrivial and trivial cases with invariant 1 and 0, respectively. The four red circles A, B, C, and D indicate the parameters for demonstrating the non-Abelian  transformations.  (c) and (d) are the plot of the left and right  ESMs of the 1D chain, respectively, where the solid (dashed) lines plot the amplitude of the $|\uparrow\rangle$ components ($|\downarrow\rangle$ components divided by   $i$) of the corresponding wave functions, with the same parameters as in (a). Adapted from Ref. \cite{caojy}.}
\label{fignas1d}
\end{figure}

Our first example is a 1D spin-1/2 lattice model which can support non-Abelian quantum operations,  the considered Hamiltonian is \cite{caojy}
\begin{equation}
\begin{aligned}
H&=\sum_l t_z(c_{l,\uparrow}^\dag c_{l+1,\uparrow}-c_{l,\downarrow}^\dag c_{l+1,\downarrow})+\text{H.c.}\\
&+\sum_l h_z(c_{l,\uparrow}^\dag c_{l,\uparrow}-c_{l,\downarrow}^\dag c_{l,\downarrow})\\
&-\sum_l i\Delta_0  (c_{l,\uparrow}^\dag c_{l+1,\downarrow}-c_{l+1,\uparrow}^\dag c_{l,\downarrow})+\text{H.c.},
\label{eq1dpolariton}
\end{aligned}
\end{equation}
which relates to Eq. (\ref{eq.simu}) according to the following: $t_{0,l\epsilon \epsilon}=t_z$, $\varphi_{l\epsilon \epsilon}=0$; $m=h_z$; $t_{0,l\epsilon\bar{\epsilon}}=\Delta_0$ and $\varphi_{l\epsilon\bar{\epsilon}}=-\pi/2$.  This Hamiltonian has a chiral symmetry and the topological invariant  is related to the  non-Abelian charge, and thus obeys the non-Abelian statistics. In Fig. \ref{fignas1d}(a), the eigenenergies of a chain with 16 lattices is plotted, where two zero-energy ESMs appear, in the middle of the energy gap, which localize at different edges. Moreover, a topological quantum phase transition critical value is $h_z=\pm 2t_z$,   where the gap closes or opens. The phase diagram of the 1D chain model with arbitrary $\Delta_0$ is plotted in Fig. \ref{fignas1d}(b), which indicates that a finite size system is sufficient to demonstrate the non-Abelian nature of the quantum operations with the prescribed parameters. Specifically, the circles A, B, C, and D in \ref{fignas1d}(b) indicate the parameters for this demonstration. The red arrow represents that $\hat{O}_1$ is executed before $\hat{O}_2$ follows; the blue arrow means that  $\hat{O}_1$ is executed after  $\hat{O}_2$, see Ref. \cite{caojy} for details. In Fig. \ref{fignas1d} (c) and (d), the two ESMs are plotted with the same parameters as in (a).
	
\begin{figure}[tbp]
\centering
\includegraphics[width=8cm]{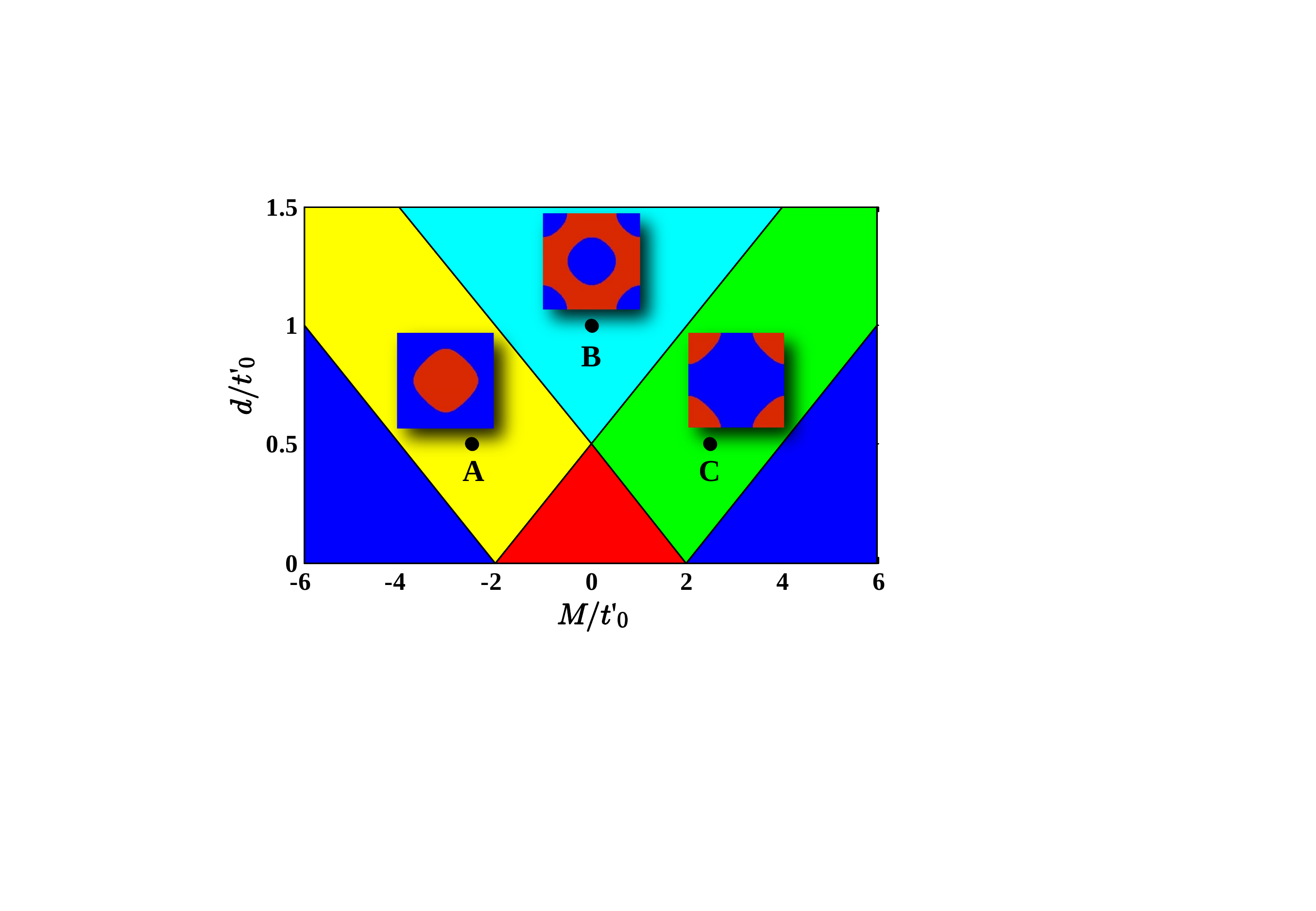}
\caption{The quantum phase diagram of a 3D nodal-loop semimetal model with the associated configurations of the winding number, with each color denotes a different phase. The yellow and green regions are topological nontrivial and one nodal loop will appear in the $k_y$-$k_z$ plane; while two nodal loops will appear for the light blue region. Their winding number configurations for points A (-2.5, 0.5), B (0, 1) and C (2.5, 0.5) of these three regions are plotted  in the three corresponding insets, where the red (blue) regions of the inserts represent the winding number being $w=1$ ($w=0$). The red region is  a topological gapped phase with $w=1$, but without nodal loops. The two blue regions are both topological trivial gapped phases. Adapted from Ref. \cite{xueso}. }
\label{fig3dphase}
\end{figure}

Our second example is to simulate a 3D nodal-loop model Hamiltonian with our 1D system with the other 2D being the parametric dimensions \cite{xueso}. The implementation is further simplified by making a unitary transformation to the simulated Hamiltonian in Eq.~(\ref{eq.simu}), i.e.,  $V=\sum_{l=1}^{N} (-i)^{l-1} \bm{I}_l$ with $\bm{I}_l=|$$\uparrow$$\rangle_l\langle $$\uparrow$$ |+ |$$\downarrow$$\rangle_l\langle $$\downarrow$$|$. When appropriately  set the parameters of the JC Hamiltonian, the target coupling can be implemented using driving with only two tunes \cite{xueso}. Then, the transformed  Hamiltonian is
\begin{align}	\label{H_nodal}
H_\text{{nod}}=&V^\dag H_\text{{ori}} V \notag\\
=&\sum_{l=1}^{N} 	m'(k_y, k_z)\bm{S^z}_{l}
+\sum_{l=1}^{N-1}\sum_{\alpha, \alpha'} it'_{0}\hat{c}^\dagger_{l,\uparrow} \hat{c}_{l+1,\uparrow} \\
&-it'_{0}\hat{c}^\dagger_{l,\downarrow} \hat{c}_{l+1,\downarrow}
+it'_{0}\hat{c}^\dagger_{l,\uparrow} \hat{c}_{l+1,\downarrow}
-it'_{0}\hat{c}^\dagger_{l,\downarrow} \hat{c}_{l+1,\uparrow} +\text{h.c.},\notag
\end{align}
where $m'(k_y, k_z)=M+2d(\cos k_y+\cos k_z)$ with $M$ being the effective Zeeman splitting energy and $d$ is the effective hopping energy along the $y$ and $z$ directions. This accomplishes the quantum simulation of  the topological nodal-loop semimetals, and the phase diagram is plotted in Fig. \ref{fig3dphase}, where the topological features can be detected using the chiral displacement operator method, see Ref. \cite{xueso} for details.

\subsection{TBT in 2D}
\begin{figure}[tb]
\center
\includegraphics[width=8.5cm]{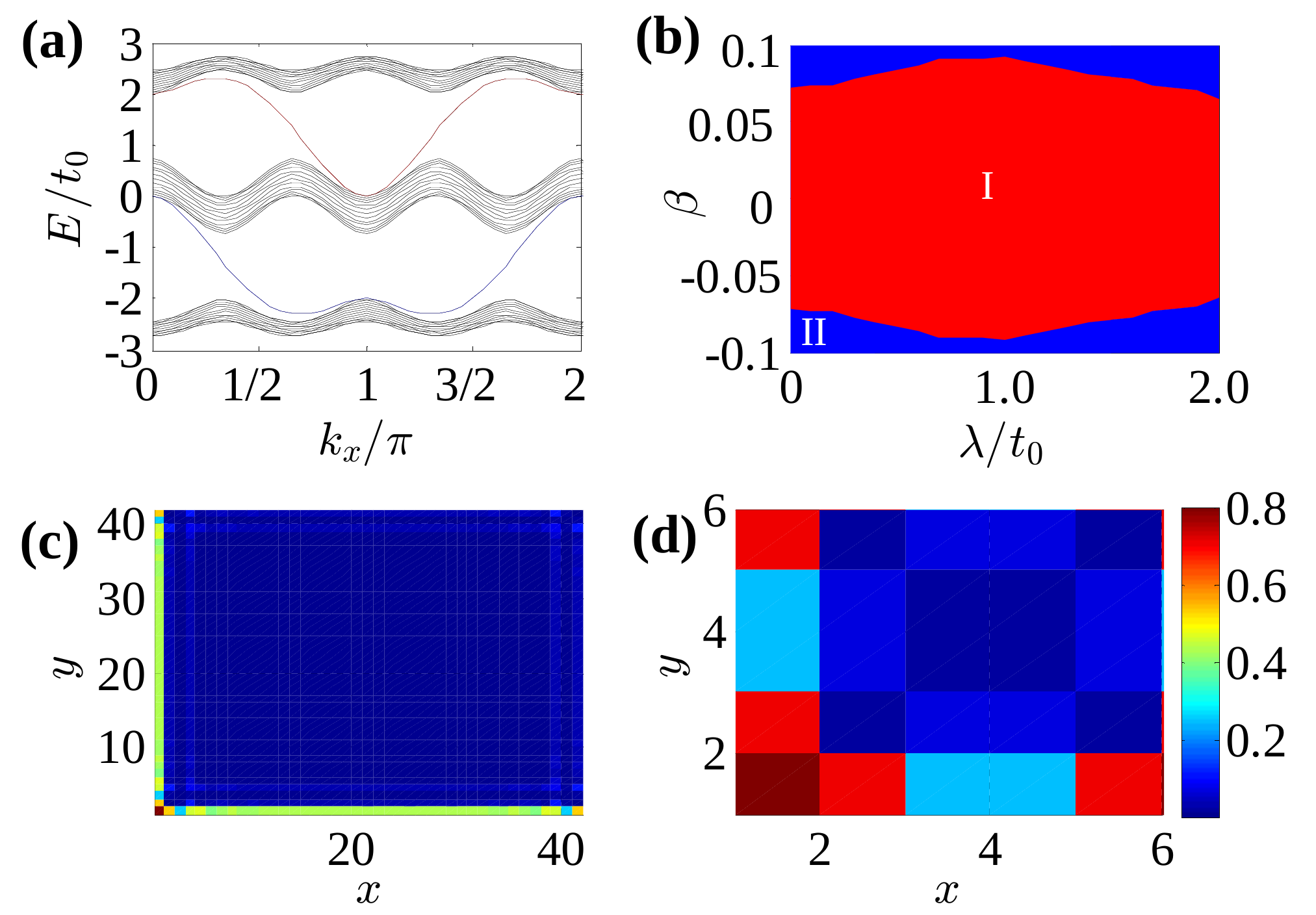}
\caption{Simulation of the quantum spin Hall effect. (a) The energy band for $\chi=0$  and $k=0$. (b)  The phase diagram  with Fermi energy in the range of $[t_0, 2t_0$], which consists of two phases, i.e., (I) quantum spin Hall phase and (II) the metal phase. The distributions of the ESMs' wave functions for (c) 42$\times$42  and (d)  6$\times$6 lattices. Adapted from Ref. \cite{liujia}. } \label{fig2dqshe}
\end{figure}

The simulation of 1D lattice Hamiltonian in Eq.~(\ref{eq.simu}) can be extended to a 2D square lattice case, where  quantum spin Hall effect can be simulated \cite{liujia}.
Then, we consider a 2D square lattice model  with the Hamiltonian of \cite{Hamiltonianqshe}
\begin{eqnarray}\label{eq2dqshe}
\mathcal{H}&=&-t_0\sum_{m,n} \left({\bf{c}}_{m+1,n}^\dag e^{i\hat{\theta}_x}{\bf{c}}_{m,n}+{\bf{c}}_{m,n+1}^\dag e^{i\hat{\theta}_y}{\bf{c}}_{m,n}+\text{H.c.}\right)\notag\\
&&+\sum_{m,n}\chi_{m,n}{\bf{c}}_{m,n}^\dag {\bf{c}}_{m,n},
\end{eqnarray}
where $t_0$ is the  neighboring hopping strength;  ${\bf{c}}_{m,n}=(c_{m,n,\uparrow}, c_{m,n,\downarrow})^T$ is a 2-component operator for a lattice site ($x=ma, y=nb$) with $a$ and $b$ are the lattice spacings and \emph{m} and \emph{n} being integers; $\hat{\theta}_x=2\pi j y\sigma_z$ and $\hat{\theta}_y=2\pi k \sigma_x$ with $(j, k)$ being parameters determined by the synthetic magnetic flux and spin mixing; $\chi_{m,n}=(-1)^n\chi$  is the on-site potential, which is staggered in \emph{y}-direction. The Hamiltonian in Eq.~(\ref{eq2dqshe}) reserves the time reversal symmetry, so it belongs to the $\mathbb{Z}_2$ class   topological phase, and it can simulate the  quantum spin Hall phase, see Fig.  \ref{fig2dqshe} for numerical results of this quantum simulation,  and Ref. \cite{liujia} for details.

\section{Experimental realizations of TBT}

The essential advantage of the descirbed parametric coupling formalisms is that they can establish the coupling between SQC elements in a time- and site- resolved way and offer consequently unprecedented flexibility to the quantum simulation of various many-body models. In recent years there have been many experimental progresses in this direction, and it is the purpose of this section to briefly introduce them.

For the formalism of coupling superconducting transmission line resonators by grounding SQUIDs, the first experiment in this direction is reported  in Ref. \cite{RoushanChiral2017NP}, where three superconducting qubits are connected to form a closed loop geometry. The coupling between the three qubits is established by grounding inductive couplers with tunable inductance, and the synthetic magnetic field penetrated in the loop is obtained from the oscillation phase of the temporal modulation of the three couplers which compensate the frequency mismatch between the qubits. From this point of view, this technique falls within the parametric coupling mechanism we have introduced in Sec. \ref{Sec Para}.

\begin{figure}[tbp]
\center
\includegraphics[width=8cm]{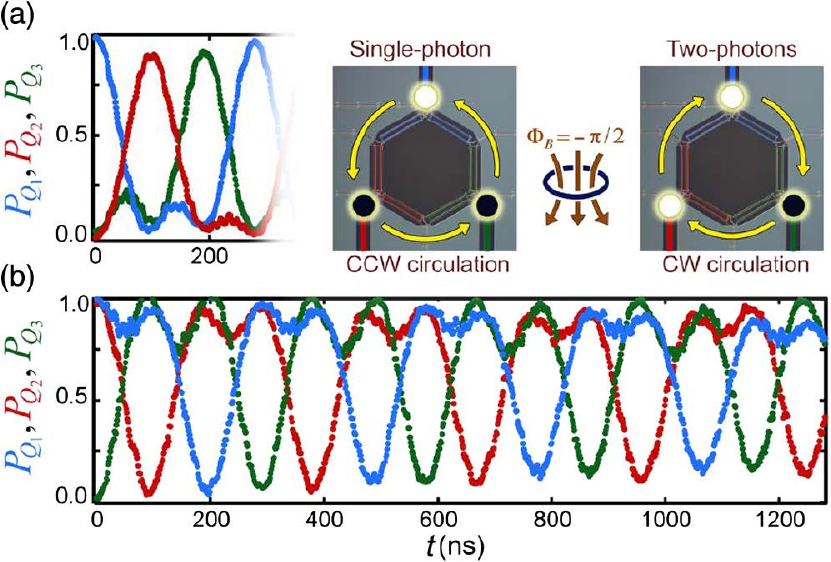}
\caption{Experimental realization of artificial magnetic field in a parametric coupled qubit necklace. The circuit and the circulation dynamics of the single- and two-excitation are sketched in the right of (a), and  the time evolution of the excitation probability
in the three qubits for the single- and two-excitation at magnetic flux $-\pi/2$ are depicted in the left of (a) and in (b), respectively. Adapted from \cite{RoushanChiral2017NP}.} \label{Fig Roushan}
\end{figure}

Due to the long coherence time currently achieved by SQC,  experiments in Ref. \cite{RoushanChiral2017NP}  are performed by first preparing the loop in a suitable one- or two-photon Fock state and then following its dynamics without any pump during the evolution. As illustrated in Fig. \ref{Fig Roushan}, the existence of the synthetic magnetic field is manifested by the directional circulation of photons in the loop with the circulation direction determined simply by the synthetic magnetic field. When the strong nonlinearity of the superconducting qubits is taken into consideration, two interacting photons would circulate in the direction opposite to that of the single photon. This can be explained by the additional $\pi$ phase induced by the Jordan-Wigner transformation. In addition, a few-body interacting ground state can also be prepared by adiabatically tuning up the synthetic magnetic field to the desired value. The same platform has been further exploited to observe the Hofstadter butterfly of one-photon states and the localization effects in the two-photon sector \cite{RoushanChiral2018Science}.

\begin{figure}[tbp]
\begin{center}
\includegraphics[width=0.45\textwidth]{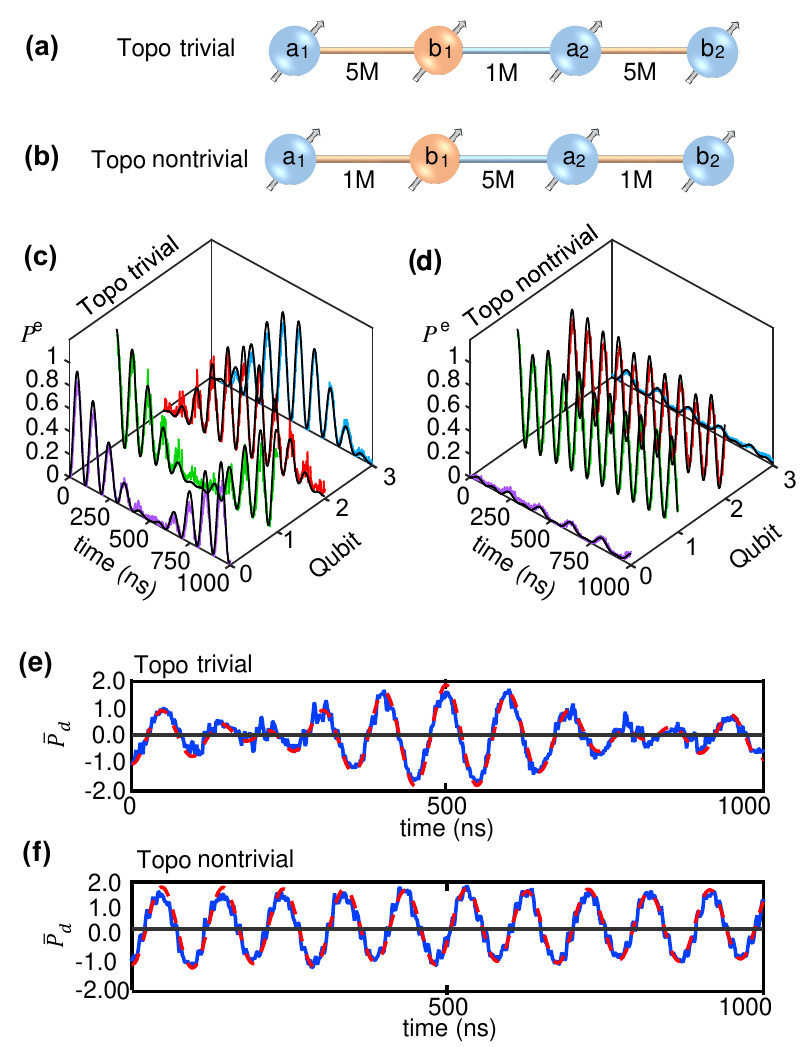}
\end{center}
\caption{Measuring the topological winding number in a four qubits chain with (a) $g_A=5$ MHz and $g_B=1$ MHz   and (b) $g_A=1$ MHz and $g_B=5$ MHz, for the  topologically trivial and nontrivial cases, respectively. (c), (d) The four qubits excitation ($P^e_{j-1}=|e\rangle_{j-1}\langle e|$) dynamics with an initial state $|gegg\rangle$  for the two prescribed coupling configurations, where dots represent the experimental data while solid lines are plotted by numerically simulations from  Hamiltonian in Eq. (\ref{eq:H}).  (e), (f) The  time-dependent average of the chiral displacement operator $\overline{P}_d =(P^e_0- P^e_2) + 2(P^e_1- P^e_3)$ for the two cases. In the long time limit, the winding number equals twice of the time-averaged $\overline{P}_d$ \cite{CD}, and thus can be extracted. Dots are experimental record, red dashed lines are from numerical simulations with the black horizontal lines are the corresponding oscillation centers, i.e., the time-averaged $\overline{P}_d$. Both experiments agree  well with the numerical simulation and theoretical prediction. The  measured topological winding numbers are 0.030 and 0.718 for the two cases, respectively. Adapted from Ref.   \cite{SSHsimu}.
} \label{figSSHdetection}
\end{figure}

On the other hand, with the tunable coupling among qubits in Eq. (\ref{eq:H}) and its demonstration  \cite{paracouple2}, we are readily in the position to simulate topological quantum phases on superconducting quantum circuits. The first example is the  1D SSH model with the Hamiltonian  of  Eq. (\ref{Eqn SSHHamiltonian}), introduced in Section \ref{SSH} and experimentally demonstrated in Ref.  \cite{SSHsimu}  by parametrically tuning the inter-qubit coupling strength in Eq. (\ref{eq:H}) in the required dimerized  form. Note that, although the Hamiltonian in Eq. (\ref{eq:H}) is of the bosonic nature, in the single excitation subspace, its topology is equivalent to that of the SSH model, which has two different topological phases with  different winding numbers.  Specifically, when the qubit couplings are tuned into $g_A < g_B$ ($g_A > g_B$) configuration, with $g_A \equiv g'_{2j+1}$ and $g_B \equiv g'_{2j+2}$  in Eq. (\ref{eq:H}), the winding number will be 0 (1), and then the system is in a topologically trivial (nontrivial) state. By monitoring the quantum dynamics of a single-qubit excitation in the chain,  the associated topological winding number can be inferred, an experimental verification of which for a four qubits case is presented in Fig. \ref{figSSHdetection}.  The measured value of the topologically trivial configuration is very close to the ideal one while the deviation from the idea value 1 in the topologically nontrivial case is mainly due to the decoherence effect of the qubits. Similarly, one can also detect the topological magnon ESMs and defect states \cite{SSHsimu}.

\section{Conclusion and prospects }
In summary, in this review we have illustrated the application of the parametric coupling method in investigating topological photonics in SQC system. Such method is a very powerful tool in the sense that it can synthesize artificial gauge field for the microwave photons with fully \textit{in situ} tunability. Using the adiabatic pumping process, the simulated topologically-protected effects can be  probed and further explored in new ways.

Despite the  rapid progress in recent years, it is our feeling that we are still in the beginning rather than the end of this research direction. Due to the flexibility of SQC system, we can expect that the PFC architecture allows the further incorporations of many other mechanisms, including on-site Hubbard interaction, disorder, and non-Hermicity \cite{Ashida}, and their interplay will definitely bring us into the realm of more fruitful physics. The first of our perspectives is that we should pay extensive attention to the advances in the technical innovation of SQC which may have potential application in quantum simulation. In particular, the reduction of the fabrication error of SQC elements can be exploited to suppress or introduce the mechanism of disorder in a controllable way. In addition, we can notice that the lattice models we discuss in this review are typically 1D or 2D. We expect that the advance of technology can  overcome such tyranny of planar connectivity which places  limit on the achievable graph configuration of the lattice models. Another expectation comes from the notation that the discussion in this review are still limited in the single particle region. Based on the fact that the synthetic gauge field has been implemented and the nonlinearity can be incorporated into SQC in a variety of manners, an immediate challenge in this field is to realize the bosonic analogues of the fractional QHE.  Our third perspective comes from the fact that the decoherence of SQC can now be suppressed to a very low level. This technique can be used to incorporate and manipulate the gain and loss of the system. From this point of view, SQC can become a versatile platform of investigating non-Hermitian topology \cite{Ashida}. Currently reported experiments of non-Hermitian topology focus mainly on the single-particle non-Hermitian skin effect in 1D lattices induced by asymmetric hopping. We then expect SQC can play an important role in this direction by pushing the research into the region of higher dimension by taking into the account of the competition of various mechanisms including interaction, gauge field, and disorder. In addition to fundamental interests, the novel properties in topological photonics are also promising in future applications. Similar to the electronic systems, a first possible application of this newly discovered topological degree of freedom should be the invention of disorder-immune devices for high-speed information transfer and processing.

\acknowledgements
We thanks Dr J. Liu for proofreading of the manuscript.

This work was supported by
the National Natural Science Foundation of China (No. 11874156 and No. 11774114),
and the Science and Technology Program of Guangzhou (No. 2019050001).

\end{document}